\documentclass[aps,pra,pifont,citesort,twocolumn,longbibliography]{revtex4-1}
\usepackage{lineno}
\usepackage{color,soul,bm}
 \usepackage{setspace}
\usepackage[utf8]{inputenc}
\usepackage{graphicx,amsmath,amssymb}
\usepackage{dcolumn}
\usepackage{bm}
\usepackage{times}
\usepackage{bbm}
\usepackage{hyperref}
\hypersetup{
    colorlinks=true,
    linkcolor=blue,
    filecolor=magenta,      
    urlcolor=cyan,
}

\newcommand{\bk}[1]{\left ( #1\right )}

\newcommand{\eqn}[1]{\begin{eqnarray} \newline #1 \end{eqnarray}}
\newcommand{\ee}{&=&}
\newcommand{\num}[1]{\begin{enumerate} #1 \end{enumerate}}
\newcommand{\hs}{\hspace{0.2cm}}

\newcommand{\bra}[1]{\left \langle#1 \right |}
\newcommand{\ket}[1]{\left |#1\right \rangle}

\newcommand{\EV}[1]{\left < #1 \right >}

\newcommand{\nn}{\nonumber}
\newcommand{\half}{\frac{1}{2}}
\newcommand{\hmin}{H_{\mathrm{min}}}
\newcommand{\hmax}{H_{\mathrm{max}}}

\newcommand{\mat}[4]{
\left(\begin{array}{cc}
#1 & #2 \\
#3 & #4
\end{array}\right)}

\newcommand{\matthree}[3]{
\left(\begin{array}{ccc}
#1  \\
#2\\
#3 
\end{array}\right)}
\newcommand{\matfour}[4]{
\left(\begin{array}{cccc}
#1  \\
#2\\
#3 \\
#4 
\end{array}\right)}

\definecolor{darkgreen}{rgb}{0.0, 0.42, 0.24}

\newcommand*{\qed}{\null\nobreak\hfill\ensuremath{\square}}%
\usepackage{epstopdf} 
\usepackage{epsfig}

\usepackage{amsmath}
\usepackage{tikz}
\usetikzlibrary{circuits.ee.IEC}

\newtheorem{Definition}{Definition}    
\newtheorem{Theorem}{Theorem}
\newtheorem{Protocol}{Protocol} 
\newtheorem{Lemma}{Lemma} 

\begin{document}

\title{Sharing classical secrets with continuous-variable entanglement: \\
Composable security and network coding advantage}
\author{Nathan Walk${}^{1}$}
\email{nathan.walk@gmail.com}
\author{Jens Eisert${}^{1,2,3}$}


\affiliation{${}^{1}$Dahlem Center for Complex Quantum Systems, Freie Universit{\"a}t Berlin, 14195 Berlin, Germany \\
${}^{2}$Helmholtz-Zentrum Berlin f{\"u}r Materialien und Energie, 14109 Berlin, Germany \\
${}^{3}$Department of Mathematics and Computer Science, Freie Universit{\"a}t Berlin, 14195 Berlin, Germany}

\date{\today}

\begin{abstract}
Secret sharing is a multi-party cryptographic primitive that can be applied to a network of partially distrustful parties for encrypting data that is both sensitive (it must remain secure) and important (it must not be lost or destroyed). When sharing classical secrets (as opposed to quantum states), one can distinguish between protocols that leverage bi-partite quantum key distribution (QKD) and those that exploit multi-partite entanglement. The latter class are known to be vulnerable to so-called participant attacks and, while progress has been made recently, there is currently no analysis that quantifies their performance in the composable, finite-size regime which has become the gold standard for QKD security. Given this -- and the fact that distributing multi-partite entanglement is typically challenging -- one might well ask: Is there any virtue in pursuing multi-partite entanglement based schemes? Here, we answer this question in the affirmative for a class of secret sharing protocols based on continuous variable graph states. We establish security in a composable framework and identify a network topology, specifically a bottleneck network of lossy channels, and parameter regimes within the reach of present day experiments for which a multi-partite scheme outperforms the corresponding QKD based method in the asymptotic and finite-size setting. Finally, we establish experimental parameters where the multi-partite schemes outperform any possible QKD based protocol. This one of the first concrete compelling examples of multi-partite entangled resources achieving a genuine advantage over point-to-point protocols for quantum communication and represents a rigorous, operational benchmark to assess the usefulness of such resources. 
\end{abstract}

\maketitle

\section{Introduction}
The desire to reliably store important information seems at odds with the desire to keep that information secret. A reasonable strategy to achieve reliability would be to generate many redundant copies of the information. However, this strategy clearly increases danger of a security breach as each copy is a new target for unauthorised access. An elegant solution to this quandary is given by secret sharing. These are protocols in which the secret is divided into pieces or {\it shares} by a dealer and distributed amongst several players such that some {\it authorised} subsets can perfectly reconstruct the secret but all other subsets gain no information whatsoever. The set of authorised subsets of a scheme is called the {\it access structure}. 

In such a scheme, any unauthorised set of 
shares may be destroyed without the secret being lost and any unauthorised set may be hacked without any information being leaked. Secret sharing can be used in many practical situations to ensure that only a sufficiently large collection of agents can authorise some action, with examples ranging from approving an expense account to ordering a military strike. Other applications include managing cryptographic keys, decentralised voting and as a primitive for secure multi-party computation. In the most common form of access structure, the dealer selects a threshold size for authorised subsets. An {\it $(n,k)$-threshold} scheme involves $n$ players of which any $k$ players can collaborate to reconstruct the secret, whilst any $(k-1)$ subset remains totally ignorant.

The concept of secret sharing has been independently conceived in a classical setting by Blakley \cite{Blakley:1979dy} and Shamir \cite{Shamir:1979cr}. However, these schemes assume that the only information received by any player is their intended share and thus they cannot be proved secure against the possibility that members of an unauthorised set are eavesdropping upon an authorised set. This problem can be solved using techniques from quantum cryptography. In the first place one could simply establish  \emph{quantum key distribution} (QKD) links between the dealer and each player in parallel \cite{PirandolaReview}. Once secret key has been established then Shamir's scheme can be safely implemented \cite{Hillery:1999tb}. The security of such schemes then follows immediately from QKD security proofs, and such schemes have already seen experimental implementation \cite{Fujiwara:2016is,Richter:2021gt}. We will refer to these schemes of \emph{bi-partite quantum secret sharing} as \emph{bQSS}. 

An alternative method, due to Hillery, Buzek and Berthiaume, is for the dealer (Alice) to create a multi-partite entangled state distributed amongst the players \cite{Hillery:1999tb}. This proposal utilised GHZ states to implement an $(n,n)$-threshold scheme and an extensive body of follow up work has since appeared \cite{Karlsson:1999ez,Qin:2007fv,Xiao:2004fw,Chen:2007ul}. A particularly interesting variant has been the work of Schmid et al.\ which does not require multi-partite entanglement, but instead transmits a quantum state between all participants who each perform a random operation \cite{Schmid:2005hz}. These should not be confused with the protocols, sometimes called quantum {\it state} sharing, where the secret to be shared is a quantum state \cite{Cleve:tg,Gottesman:2000vx}. Crucially, almost immediately following the original HBB paper, it has been pointed out that these protocols are vulnerable to so-called \emph{participant attacks} \cite{Karlsson:1999ez,Qin:2007fv} and the security of these schemes could not be rigorously established.

Subsequently, several works \cite{Markham:2008ta,Keet:2010cf,Marin:2013dc,Lau:2013gm} have identified \emph{graph states} 
\cite{Hein04}
as a valuable resource for secret sharing (with classical and quantum secrets) which allow for more general $(n,k)$-threshold schemes and highlight an elegant connection between secret sharing and error correction codes. This setting is
conceptually interesting. At the same time, it has
become more technologically plausible. Substantial theoretical progress has also been made on how to distribute graph states
in multi-partite quantum networks 
\cite{HahnPappaEisert,Wehner18}.
Whilst these proposals
have comprehensively answered the questions of how secrets can be successfully reconstructed by the authorised subsets, the security analysis against dishonest parties remained unsatisfactory because the problem of participant attacks remained unsolved. 

In contrast to QKD where the dishonest party is completely shut out of the parameter estimation process, secret sharing typically includes all players in the certification procedure. This opens up loopholes regarding the order in which information (measurement bases and outcomes) is announced that can be exploited by dishonest players to avoid detection. Thus, while many experimental implementations have appeared \cite{Tittel:2001gn,Lance:2004do,Chen:2005fs,Schmid:2005hz,Gaertner:2007tt,Bogdanski:2009jo,Bell:2014ez,Grice:2015jw,Fu:2015df,Armstrong:2015he,Cai:2017cp,Zhou:2018hg,DeOliveira:2019ta}, they have all only been analysed either under various assumptions (e.g., perfect state transmission, asymptotically many rounds or in some cases specific eavesdropping strategies) or restrictions upon the players and the eavesdropper and none were shown to be secure against arbitrary participant attacks. We note that some works on sharing entangled quantum states do rigorously address the participant attack \cite{Markham:2018uw,Pappa:2012p5495}, but only by utilising a pre-existing secret sharing protocol for classical strings. This is reasonable when the goal is to leverage the security of classical bQSS to ultimately share a quantum state, but would be redundant for sharing classical secrets which is our primary concern here.

The problem has finally been resolved, at least in the asymptotic limit of infinitely long key exchange, first by Kogias et al. \cite{Kogias:2017jz} in the context of \emph{continuous-variable (CV)} graph states 
\cite{PhysRevA.76.032321,PhysRevLett.97.110501,PhysRevA.79.062318,Ohliger:2010kv}
and later by Williams et al.\ \cite{Williams:2019kb} for \emph{discrete-variable (DV)} GHZ states, where the latter also has carried out a proof-of-principle demonstration. Using different methods, both works manage to reduce the problem to essentially a minimisation over bi-partite scenarios where tools from QKD analysis can be applied, but without leaving any room for participant attacks. Follow up works has extended \cite{Williams:2019kb} to the CV regime \cite{Grice:2019ff} and included a finite statistical-analysis under the assumption of Gaussian collective attacks \cite{Wu:2020bs}. Importantly, none of these works give any instances where a genuinely 
\emph{multi-partite approach} results in any improvement in performance, in fact in Ref.\ \cite{Kogias:2017jz} it is shown that their multi-partite entangled protocol is strictly inferior to bQSS over the networks they consider. 

This gives rise to a most pressing situation: The vision of 
\emph{quantum
networks} \cite{Kimble:2008p5456,WehnerQuantumInternet,Meter:2012p5453,TeleportationReview}, with notions of a 
\emph{quantum internet} in mind, seems to suggest that a wealth of new multi-partite protocols based on multi-partite entanglement opens up. Yet, at the same time it 
seems excessively difficult to identify schemes that actually 
obtain an advantage based on the availability
of multi-partite entangled states
under realistic conditions. This obstacle is largely overcome here.

In the following, we will first explain the differing analyses of Refs.\ \cite{Kogias:2017jz} and \cite{Williams:2019kb}, and quantitatively improve upon the rates calculated in the former work. We then lift the analysis to consider arbitrary attacks in the composable, finite-size regime. \emph{Composable security} is a 
particularly stringent notion of security in which the protocol remains secure even if arbitrarily composed with other instances of the same or other protocols. To be in the \emph{finite-size regime} also seems a practical necessity given that asymptotic settings usually
refer to extremely long sequences of key exchange. 

Finally, we turn to the main contribution of this paper, which is to evaluate performance over bottleneck networks of 
lossy channels and demonstrate a genuine quantitative advantage for protocols exploiting multi-partite entanglement in CV graph states. In the limit of asymptotic key rates, we show an unconditional advantage, in the sense of outperforming the so-called \emph{PLOB bound} which represents the ultimate limit on point-to-point QKD protocols \cite{Pirandola:2017jk}. For large but finite squeezing the multi-partite scheme can outperform the PLOB bound for a transmission radius of up to 4km of optical fibre. We model a realistic multi-partite experiment and find that even in the composable, finite-size regime an advantage exists over a CVQKD scheme with the same resources. This represents -- once again -- a rare concrete example of a multi-partite entanglement advantage for quantum cryptography over realistic networks.

\section{Security of secret sharing}

\begin{figure}[htb]
\includegraphics[width=0.5\textwidth]{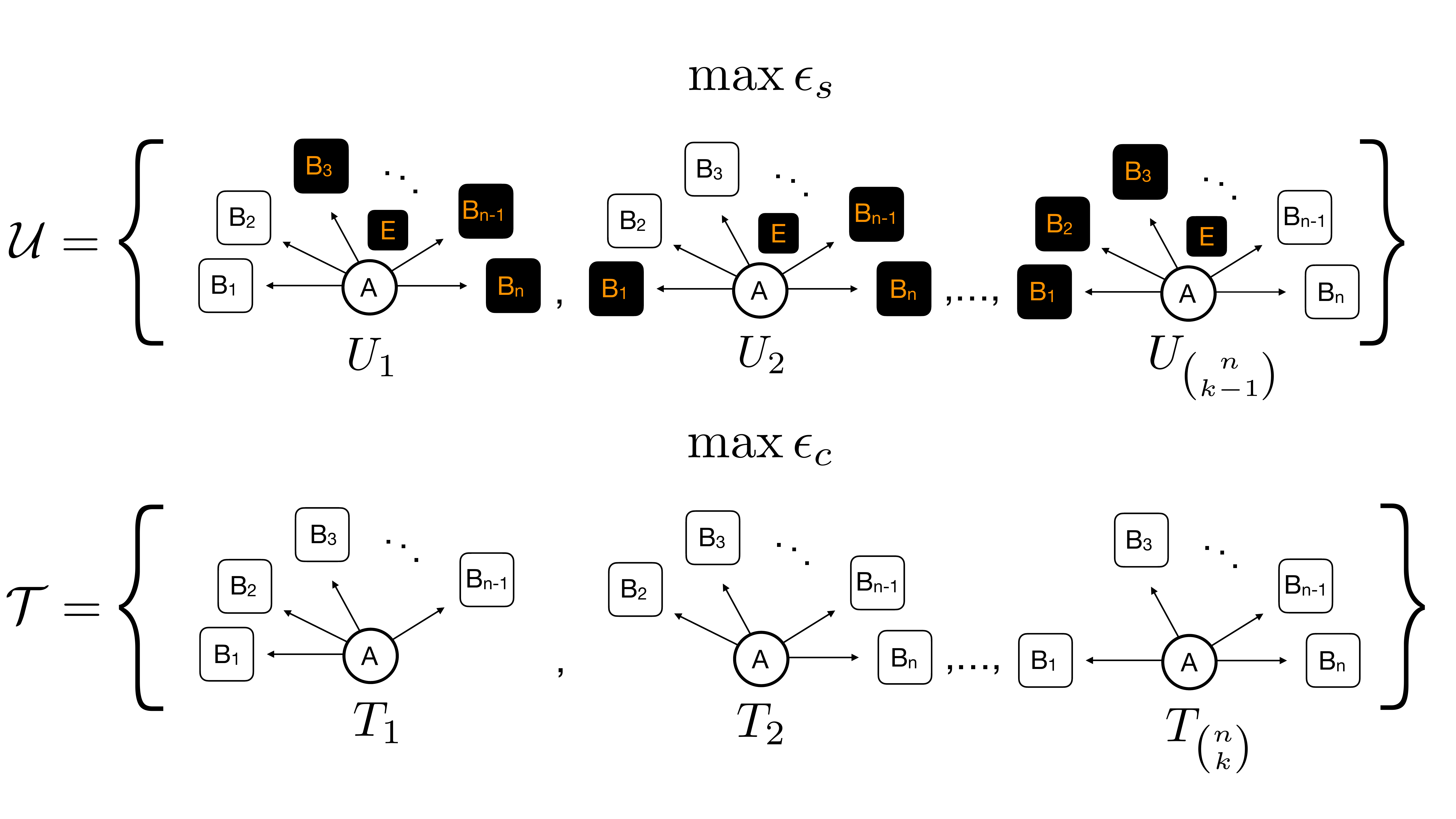}
\caption{Security analysis for an $(n,n-1)$-threshold scheme. The performance of the scheme is assessed by taking the worse case values of the failure parameters (see Definition~\ref{SSdef}) for both 
secrecy $\epsilon_s$ and the correctness $\epsilon_c$. For a general ($n,k$)-scheme, $\epsilon_s$ is maximised with respect to all for the $\binom{n}{k-1}$ possible unauthorised subsets of players who might be collaborating with the eavesdropper and $\epsilon_c$ is maximised with respect to the $\binom{n}{k}$ authorised subsets who may wish to later collaborate to reconstruct the secret. \label{nkschem}}
\end{figure}

The idea of sharing a classical secret with quantum technology is to distribute a random key that has precisely the desired access structure, and then encrypt that actual secret via a one-time pad encoding. In fact, like standard QKD, this protocol technically carries out key expansion rather than distribution since a small amount of pre-shared key must already exist to authenticate any public communication and to carry out privacy amplification. Consider an $(n,k)$-threshold scheme where, in each round of the protocol, a multi-partite entangled state is shared between $n$ players ($B_1,\dots, B_n$) and a dealer, Alice, who measures her part of the state in one of two conjugate bases. Measurements in one basis will be used to form a secret key while the others will be publicly disclosed and used for certification. Typically this is done asymmetrically with $p$, the probability for a certification round, satisfying $p\leq \half$. To process her measurement outcomes into a secret key with the desired access structure, Alice must determine two parameters: On the one hand, this is the amount of privacy amplification required such that the key appears random to any $(k-1)$-subset who might be in league with the eavesdropper. On the other hand, this is the amount of error reconciliation information she must transmit to ensure any authorised $k$-party subset can reconstruct the secret key. To this end, 
we need to define the following sets: The set of all players (Bobs) \begin{equation}
 \mathcal{B} = \{B_1,B_2,\dots ,B_n\};
    \end{equation}
the set of all authorised or trusted subsets of $k$ players $\mathcal{T} = \{T_1,T_2,\dots ,T_{\binom{n}{k}}$ where, e.g., 
\begin{equation}
T_1 = \{B_1,B_2,\dots ,B_k\} 
   \end{equation}
   and so on; the set of all unauthorised or untrusted subsets of $(k-1)$ players $\mathcal{U} = \{U_1,U_2,\dots ,U_{\binom{n}{k-1}} \}$ where, e.g., 
  \begin{equation} 
   U_1 = \{B_1,B_2,\dots ,B_{k-1}\} 
     \end{equation}
     and so on. To determine the extractable key, Alice must take worst case estimates for the secrecy over the $\binom{n}{k-1}$ unauthorised subsets and for the correctness over the $\binom{n}{k}$ authorised subsets (Fig.~\ref{nkschem}). 

Moreover, Alice must do this in a way that prevents any \emph{participant attacks}, which typically exploit the order in which certification information (measurement bases and outcomes) is announced by the players. 
This is the critical point where the security of most previous multi-partite schemes can be completely broken.
Note that in QKD protocols, the measurement bases can in principle be established beforehand for an $L$ round scheme at the cost of $\sim L h_2(p)$ extra bits of pre-shared key where $h_2$ is the binary entropy function. However, such a scheme is \emph{a priori} forbidden for a secret sharing scheme as it is crucial that the potentially dishonest players do not know ahead of time which rounds will be used for certification.

For example, the original HBB  protocol attempts to certify a GHZ-state scheme by having the players randomly switch between measuring in the Pauli $X$ or $Y$ basis and then verifying that measurement combinations corresponding to GHZ stabilisers behave as expected \cite{Hillery:1999tb}. However, if measurement bases are announced first and a dishonest player (Bob) knows that he will be the last to make an announcement, he can cheat perfectly as follows \cite{Karlsson:1999ez}. In the transmission phase Bob intercepts all of the GHZ photons sent by Alice and instead establishes bi-partite maximally entangled states between himself and the other players. Once all other players announce their measurement basis, Bob measures his maximally entangled pairs and also immediately learns all other players measurement outcomes. Finally, he can use his knowledge announced bases to ensure the round is only kept if Alice measures in a basis of Bob's choosing. For example, if all other players announce the $X$ basis, Bob knows if he also announces $X$ then the round will only be kept if Alice also measures $X$ since only that choice corresponds to a GHZ stabiliser. Bob now measures his intercepted GHZ particles and perfectly learns Alice's X measurement outcome and, along with his knowledge of the other players outcomes, can calculate what he should announce in order to perfectly recreate the GHZ correlations. Thus, the scheme is completely broken but Bob remains undetected. Other attacks are also possible if measurement outcomes rather than bases are announced first \cite{Karlsson:1999ez}.

Two solutions to this problem have emerged. One strategy is to test each potential unauthorised subset by simply excluding all players in that subset from the certification process \cite{Kogias:2017jz}. The other is to have the dealer randomly select an unauthorised subset that is included in the certification test but forced to make all their announcements first. This essentially reduces each test to a QKD protocol with an untrusted source \cite{Williams:2019kb}. One way to enforce this ordering would be to instruct members of the complementary set $C_j$ to withhold their announcements until they have received that rounds results from the corresponding $U_j$. Note that the protocol of Ref.\ \cite{Williams:2019kb} therefore comes with additional classical communication overheads.

The two strategies cannot easily be compared in general. Whilst the technique of Ref.\ \cite{Kogias:2017jz} is simpler, it will always obtain lower correlations between Alice and any given subset as it does not make use of the announced results from untrusted parties. For example, while this method has been  shown to predict positive rates for CV graph states it always results in a zero key rate if applied to the original HBB protocol. This stems from the fact that, for a GHZ state, tracing out even a single party results in completely uncorrelated noise in either the Pauli $X$ or $Y$ bases of the other participants.

On the other hand, the proof of Ref.\  \cite{Williams:2019kb} can be applied to an HBB-type protocol but requires $\binom{n}{k-1}$ different data sets for parameter estimation (one for each $U_j$ announcing first), meaning that much more data must be sacrificed for certification. Moreover, this latter protocol stipulates that the bases be chosen symmetrically (i.e., $p= 1/2$) which halves the achievable rate and it is also necessary to acquire certification measurements in both bases, further driving down performance. We will show later that these restrictions are indeed mandatory for security to be maintained. 

For a general protocol with arbitrary players and including finite-size effects it is possible that there are instances where the approach of Ref.\  \cite{Williams:2019kb} could prove superior. However, for the case we will consider in Section~\ref{bottleneck}, namely three parties utilising a bottleneck network, the reduction of the key rate by a factor of $1/2$ already precludes any possible advantage for the multi-partite entangled scheme. Therefore, we will instead use the protocol of Ref.\  \cite{Kogias:2017jz} as our starting point for the remainder of this work. Further discussion of the security, potential drawbacks and applications for the work of Williams et al.~can be found in Appendix~\ref{Williams}. 

\begin{Protocol}[Entanglement-based secret sharing\label{Prot}]
An $(n,k,m,t,p)$-protocol for entanglement-based secret sharing involves the following steps:
\num{

\item The dealer (Alice) establishes sets of pre-shared keys: a bi-partite key with each player to authenticate classical communication channels and a joint key that satisfies the intended $(n,k)$-threshold access structure as a seed for privacy amplification.
\item An $(n+1)$-partite entangled state is distributed amongst players and the dealer (Alice) through untrusted quantum channels.
\item Alice measures her part of the state in one of two non-commuting bases, $\mathbb{X}$ and $\mathbb{P}$ with probability $p$ and $1-p$ respectively. We will denote the key generation measurement $\mathbb{X}$ and the certification measurement $\mathbb{P}$. 
\item If the players are honest they also randomly choose between the measuring $\mathbb{X}$ and $\mathbb{P}$ on their systems according to the same probability. If they are dishonest, nothing is assumed about their actions at this point. We denote the $i^{th}$ authorised set of $k$ players as $T_i$, the $j^{th}$ unauthorised subset of $k-1$ players as $U_j$ and the corresponding complementary subset of $n-k+1$ players as $C_j$.
\item Following Ref.\  \cite{Kogias:2017jz}, all players announce their measurement bases for all rounds in any order. If the announced values of any of the $T_i$ or $C_j$ are consistent with Alice's measurement choice this round is kept. Depending upon the correlation measure to be employed, Alice may only designate $\mathbb{P}$ measurements for disclosure, or she may also select a random subset of her $\mathbb{X}$ basis rounds. This process is repeated until Alice has designated $m$ rounds to be used for key generation and $t$ rounds for parameter estimation. Using this single parameter estimation data-set Alice computes a correlation measure between herself and each complementary subset, $C_j$. In any given computation Alice simply ignores all announcements from any player in the corresponding unauthorised set $U_j$. If the correlations are below a certain level, the protocol aborts. Depending upon protocol specifics there may also be other checks carried out (e.g. an energy test or a decoy state analysis) which, if failed, will also cause the protocol to abort. 

\item If the test passes, this results in correlated variables $\bk{\mathbf{X}_A, \mathbf{X}_{B_i}}$ which describe the  measurements of Alice and each of the authorised subsets. Alice proceeds with error reconciliation which leaks a maximum of $\ell_\mathrm{EC}$ bits of information and privacy amplification utilising two-universal hashing. The correctness of error reconciliation is verified with a check that involves announcing a further hash of length $\log_2\epsilon_c$ bits computed with a pre-shared seed. If this check also does not abort this results in a final keys of length $l$ $\bk{\mathbf{S}_A, \mathbf{S}_{B_i}}$ for Alice and each authorised subset.
}

\end{Protocol}

We can now formally state our definitions for secret sharing in the composably secure framework established for QKD \cite{Renner:2005p464,Tomamichel:2012p7120,Portmann:2014uz}. Let $p_{\mathrm{pass}}$ be the probability that the protocol does not abort and define the joint state (conditioned on passing) between the register of Alice's final key and the $j^{\mathrm{th}}$ untrusted subset in collaboration with the eavesdropper as the classical-quantum state,
\eqn{\rho_{\bm{S}_AE} = \sum_{\bm{s}_A}p(\bm{s}_A) \ket{\bm{s}_A}\bra{\bm{s}_A}\otimes \rho_{E,U_j}^{\bm{s}_A} \label{S}}
where the sum is over all possible $l$-bit strings that could make up the key and $\rho_{E,U_j}^{\bm{s}_A} $ is the state of Eve and the $j^{\mathrm{th}}$ unauthorised subset given a certain value of the key.

\begin{Definition}[Notions of secret sharing schemes]
A secret sharing scheme as defined in Protocol~\ref{Prot} that outputs a state of the form (\ref{S}) is
\begin{itemize}
\item {\it $\epsilon_c$-correct} if 
\eqn{\max_i \left\{ \mathrm{Pr}[\bm{S}_A \neq \bm{S}_{B_i}] \right \} \leq \epsilon_c} 
and 
\item $\epsilon_s$-secret if 
\eqn{\hs \max_j \left\{p_{\mathrm{pass}} D\bk{\rho_{\bm{S}_AE}, \tau_{\bm{S}_A}\otimes\sigma_{EU_j}} \right \} \leq \epsilon_s \label{sec}}
where $D(\cdot,\cdot)$ is the trace distance and $\tau_{\bm{S}_A}$ is the uniform (i.e., maximally mixed) state over $\bm{S}_A$.
\end{itemize}
A protocol is {\it ideal} if it satisfies 
$\epsilon_c = \epsilon_s=0$, and it is called $\epsilon_{\mathrm{sec}}$-secure if  $\epsilon_{\mathrm{sec}}=\epsilon_c + \epsilon_s$. This means that there is no device or procedure that can distinguish between the actual protocol and an ideal protocol with probability higher than $\epsilon_{\mathrm{sec}}$. 
\label{SSdef}
\end{Definition}

If we define $\ell_{\mathrm{EC}}^i$ as the amount of error correction needed for the $i^{\mathrm{th}}$ authorised subset it can be shown using results from the QKD literature \cite{Tomamichel:2011ci,Tomamichel:2012p7120}, that a key of length 
\eqn{l  &=&  \hmin^\epsilon(\bm{X}_A|E,U_j) - \ell_{\mathrm{EC}}^i - \log_2\frac{1}{\epsilon_c\epsilon_1^2} +2 \label{kth1}}
that is  $\epsilon_c$-correct and $\epsilon_s$-secret against the $j^{\mathrm{th}}$ unauthorised subset where $\hmin^\epsilon(\bm{X}_A|E,U_j)$ is the conditional smooth min-entropy evaluated over the state given in (\ref{S}) and $\epsilon$ and $\epsilon_1$ are positive constants proportional to $\epsilon_s$ which can be optimised over. The necessary results have also been proven for infinite-dimensional systems which we require here \cite{Furrer:2012p8365,Furrer:2014ig}. Considering Definition~\ref{SSdef}, 
the extractable amount of key for secure secret sharing is then
\eqn{l  &=&  \min_j \hmin^\epsilon(\bm{X}_A|E,U_j) - \max_i \ell_{\mathrm{EC}}^i \nn \\
&-& \log_2\frac{1}{\epsilon_c\epsilon_1^2} +2 . \label{lss}}
A standard figure of merit for a cryptographic protocol is the secure fraction ${l}/{L}$ - the ratio of secure output bits to the number of attempted channel or network uses. 

The choice of error reconciliation code fixes $\ell^i_\mathrm{EC}$  with respect to $\epsilon_c$ so the major remaining task is lower bounding $\hmin^\epsilon(\bm{X}_A|E,U_j)$ for a given $\epsilon_s$ from the data gathered during parameter estimation. This is the crucial step where a mistake could create vulnerabilities to participant attacks. A commonly used tool for this task is an entropic uncertainty relation for the observables $\mathbb{X}_A$ and $\mathbb{P}_A$.

Without loss of generality, the overall state can taken to be pure ($\rho_{AU_j\mathcal{C}_jE} = \ket{\Psi_{AU_jC_jE}}\bra{\Psi_{AU_jC_jE}}$). In this case, it has been shown that the following \emph{entropic uncertainty relation} holds for the $m$-round state used for key generation \cite{Furrer:2012p8365,Furrer:2014ig},

\begin{equation}\hmin^\epsilon\left(\bm{X}_{A}^m | E, U_j\right)+\hmax^\epsilon\left(\bm{P}_{A}^m | C_j\right) \geqslant m\hspace{.2mm} q(\mathbb{X}_A, \mathbb{P}_A)\label{eur}
\end{equation}
where the constant $q(\mathbb{X}_A, \mathbb{P}_A)$ quantifies the complementarity of the two measurement bases and we have added superscripts to Alice's variables to emphasise that we are referring to the $m$-rounds to be used for key generation.
This result would appear to immediately solve our problem in that it can be rearranged to lower bound the quantity of interest, $\hmin^\epsilon\left(\bm{X}_{A}^m | E U_j\right)$, in terms of correlations between Alice and the trusted subset $\mathcal{C}_j$. Importantly however, relations like this are counterfactual in that they describe two hypothetical situations (Alice measuring either $\mathbb{X}_A$ or $\mathbb{P}_A$) only one of which can actually happen. Thus we do not directly have access to the correlations between $\bm{P}_A^m$ and $\mathcal{C}_j$ that appear in (\ref{eur}), as all $m$ of these rounds are in fact measured by Alice in the $\mathbb{X}_A$ basis. 

Instead, we have the strings $\bm{P}_A^{t_j}$ and $\bm{P}_{C_j}^{t_j}$ arising from the $t$ rounds announced during parameter estimation. Note that, in general, we will have $t_j<t$. This is because any one of the $t$ total parameter estimation rounds might only be useful for estimating correlations with the complementary subset $C_j$ but not with some other subset $C_k$. Crucially, provided that the parameter estimation rounds were truly selected at random, then this is a standard problem in random sampling without replacement. We can apply the result of Serfling \cite{Serfling:1974dx} to bound the correlations would have been counterfactually observed between $\bm{P}_A^m$ and $\bm{P}_{C_j}^m$, given the actually observed correlations between $\bm{P}_A^{t_j}$ and $\bm{P}_{C_j}^{t_j}$. It is then possible to bound the min-entropy \cite{Furrer:2012p8365,Furrer:2014ig}. However, this is only valid for genuinely random sampling and it is precisely this condition that is violated by the participant attacks outlined previously where parameter estimation process involves all players simultaneously, including the potentially dishonest ones. 

Recall that in the example participant attack on the HBB protocol, dishonest Bob learns the measurement bases of the other players before making his own announcement. He could then choose his announced basis to deterministically ensure that this particular round will only be kept if Alice measured in a particular basis. If the certification measurement is fixed to be $\mathbb{P}_A$, this means dishonest Bob can control the sampling procedure such that the correlations between $\bm{P}_A^{t_j}$ and $\bm{P}_{C_j}^{t_j}$  are not valid as a fair sample to estimate those between the counterfactual $\bm{P}_A^m$ and $\bm{P}_{C_j}^m$. Note that this loophole would still exist even dishonest Bob was forced to announce first if it was still the case that all certification measurements made in the $\mathbb{P}_A$ basis. This is why in the protocol of Williams et al., which includes all players in the certification step, it is mandatory that bases be chosen symmetrically and a random subset of each basis is used to certify the secrecy of the other.

In the Protocol~\ref{Prot}, whether a round is kept for any fixed value of $j$ is determined solely by the bases of Alice and the complementary set $\mathcal{C}_j$ so this problem is automatically avoided and the relation in (\ref{eur}) can be successfully utilised. Specifically, it can be shown that if a correlation measured defined for two $m$-length strings ($\bm{X},\bm{Y})$ as $d(\bm{X},\bm{Y}):= (1/m) \sum_{i=1}^m |\bm{X}-\bm{Y}|_i$ is greater than some threshold, $d_0$, then $\hmax^\epsilon\left(\bm{P}_{A}^m | C_j\right) $ can be upper bounded. Using Serfling's bound the observed correlations $d(\bm{P}_A^{t_j},\bm{P}_{C_j}^{t_j})$ can be used to estimate a $d_0$ that would have been satisfied by $d(\bm{P}_A^m,\bm{P}_{\mathcal{C}_j}^m)$. The same arguments apply for other quantities obtained during parameter estimation, such as a covariance matrix.

Up until this point, these arguments can be applied to either a DV or CV realisation of Protocol~\ref{Prot}. However, there are still several issues that need to be dealt with in order to evaluate the secure fraction for a realistic CV protocol where the conjugate bases are approximate quadrature measurements made via homodyne detection. Two primary issues are that real quadrature measurements have a finite resolution ($\delta_{\mathbb{X}},\delta_{\mathbb{P}}$) and a finite range $\bk{[-M_{\mathbb{X}},M_{\mathbb{X}}], [-M_{\mathbb{P}},M_{\mathbb{P}}]}$. The first problem can be dealt with evaluating the complementarity constant in (\ref{eur}) for a coarse-grained observable that accounts for the finite resolution and the second by introducing an additional test to the protocol where the dealer taps off a small portion of the their incoming light with a beam-splitter of transmission $\eta$ and makes an estimate of the input energy, either via heterodyne \cite{Furrer:2014ig} or direct \cite{Drahi:2020gw} detection. The protocol is aborted if too large a value is observed, which ensures that the energy of the input state is appropriate for the range of the detectors being used. Following previous CVQKD literature \cite{Furrer:2012p8365,Furrer:2014ig} shown that, given a correlation threshold $d_0^j$ passed by the set $\mathcal{C}_j$ and an energy threshold $\alpha$, an ($\epsilon_s+\epsilon_c$)-secure secret string can be extracted of length,
\eqn{l\ee m\left[q(\delta_{\mathbb{X}},\delta_{\mathbb{P}})-\max_j \log _{2}\bk{ \gamma\left(d_{0}^j+\mu\right)}\right]\nn \\
&-&\max_i \ell^i_{\mathrm{EC}} -\log _{2} \frac{1}{\epsilon_{1}^{2} \epsilon_{c}}+2 \label{secfrac}}
where 
\begin{equation}
\gamma(t)=\left(t+\sqrt{1+t^{2}}\right)\left(\frac{t}{\sqrt{1+t^{2}}-1}\right)^{t}
\end{equation}
and $\mu$ is a complicated constant that depends upon the thresholds ($d_0,\alpha$), block sizes ($m,t$), security parameters ($\epsilon_s,\epsilon_c$) and detection parameters ($\delta_{\mathbb{X},\mathbb{P}}, M_{\mathbb{X},\mathbb{P}}, \eta$). A full security proof is given in Appendix~\ref{security}. The entropic uncertainty relation in (\ref{eur}) is presently the only known technique for the composable finite-size analysis of homodyne based protocols, but it is known not to be tight in typical QKD scenarios \cite{Furrer:2014ig,Walk:2016jx,Walk} leading to overly pessimistic predictions.

We can also calculate simpler, idealised rates in the limit of infinitely many rounds and perfect detection and information reconciliation. Here, it has been shown that so-called collective attacks - where the malicious parties act in an i.i.d.\ (independent and identically distributed) manner - are optimal \cite{Renner:2009p1}. We first evaluate the secure fraction in (\ref{lss}) directly where the min-entropy limits to the von Neumann entropy via the asymptotic equi-partition theorem $\lim_{m\rightarrow \infty }\hmin^\epsilon(\bm{X}^m_A|E,U_j) = mS(X_A|E,U_j)$ where 
\begin{eqnarray}
    S(X_A|E,U_j) &=& H(X_A)+\sum_{x_A}p(x_A) S(\rho_{E,U_j}^{x_A}) \label{cvn}\\
    &-& S(E,U_j)\nonumber
\end{eqnarray}
is the \emph{conditional von Neumann entropy} of $X_A$ given the quantum system $E,U_j$ with $H(X) = -\sum_{x}p(x)\log_2p(x)$ and $S(\rho) = -\mathrm{tr}\bk{\rho\log_2\rho}$  the \emph{Shannon}
and 
\emph{von Neumann entropies}, respectively. Then, with perfect error reconciliation, we have that the amount of leaked information during reconciliation with a trusted subset becomes $\ell^i_{\mathrm{EC}} = m H(X_A|X_{T_i})$. Finally, in the asymptotic limit only a negligible amount of data needs to be sacrificed for parameter estimation so we have that $p\rightarrow 1$ and thus $m\rightarrow L$. In this limit, 
we recover the expected asymptotic formulas
\eqn{K_{\mathrm{SS}} &:=& \lim_{L\rightarrow\infty}\frac{l}{L}\nn\\
 &=& \min_j S(X_A|E,U_j) - \max_i H(X_A|X_{T_i}) \nn \\
 \ee \min_i I(X_A:X_{T_i}) - \max_j \chi(X_A:E,U_j) \label{kasymp}}
 where in the third line we have rewritten the key rate in terms of the \emph{mutual information} 
 \eqn{I(X_A:Y) := H(X) - H(X|Y)\label{iab}} and the \emph{Holevo quantity} 
 \begin{equation}
 \chi(X_A:E) := S(E) - \sum_{x_A} p(x_A)S(E|x_A). \label{holevo}
 \end{equation}
These asymptotic results have been derived in Ref.\ \cite{Kogias:2017jz}, however, 
the manner in which they go on to bound these quantities is unnecessarily pessimistic. This is because they also utilise an entropic uncertainty relation for ideal quadrature measurements following the results for \emph{one-sided device independent (1sDI)} CVQKD in Ref.\ \cite{Walk:2016jx}. The authors of Ref.\ \cite{Kogias:2017jz} go on to describe the 1sDI nature of their proof as being crucial for protection against participant attacks. However, as we have explained above the essential ingredient in their security proof is actually that there is always some part of the parameter estimation process where each possible untrusted subset is excluded. Within a given check, it is perfectly safe to assume that the trusted parties have well characterised devices and therefore a 1sDI protocol is not mandatory. Instead, in the asymptotic regime, where collective attacks are known to be optimal \cite{Renner:2009p1}, we are free to use the results from Refs.\ 
\cite{GarciaPatron:2006p381,Navascues:2006p805} based on Gaussian extremality to obtain tighter rates. Note that to apply these methods it is necessary to reconstruct an entire covariance matrix rather than just a correlation measure.

\section{Network coding advantage in bottleneck networks \label{network}}

Recent work has seen a substantial interest in
notions of network coding and multi-partite entanglement for quantum communication, aimed at understanding
in what way multi-partite schemes may
outperform point-to-point schemes. 
Indeed, important steps have been
taken, in particular on how multi-partite
states can be distributed and 
manipulated
\cite{Epping:2016hc,Epping:2017dx,HahnPappaEisert,Pivoluska:2018bp,Dahlberg:2018ku}. This is 
largely motivated by recent 
experimental and technological
developments \cite{Wang:2016dk,Proietti:2020ty} that render ideas of
\emph{quantum networks} and the quantum
internet plausible \cite{Wehner:2018cu,Kimble:2008p5456}.
At the same time, it seems less clear how to
arrive at a  setting in which
there is a genuine quantifiable
network coding advantage over 
point-to-point schemes. 

In this section, we make an affirmative claim
of a network coding advantage in a CV
bottleneck network. At the heart of the 
protocol devised is the concept of a 
\emph{CV graph state}, the continuous
variable analog of a graph state. In the canonical construction (see Appendix~\ref{CVgraphs}) each node of the graph is intialised in a squeezed vacuum state and each edge corresponds to an entangling gate that also requires active squeezing.
The first work to show a concrete performance enhancement when using multi-partite entanglement for cryptography \cite{Epping:2017dx} has focused on \emph{conference key agreement (CKA)} 
sometimes called NQKD \cite{Murta:2020de}.
In a CKA protocol all players are assumed to be honest the goal is for the dealer establish a key that can be reconstructed by each player individually. In Ref.\  \cite{Epping:2017dx}, the authors have  considered a network featuring a bottleneck where the dealer, Alice, is separated from the other players, by a central hub $H$, with the ability to carry out entangling gates. Each player is connected to $H$ by a quantum channel. 

For the case of perfect channels, a bi-partite scheme for either CKA or QSS would require $n$ network uses to conduct a QKD protocol with each Bob, but for a multi-partite entanglement based scheme only one use would be necessary. In Ref.\ \cite{Epping:2017dx} the authors analysed a GHZ state protocol and found an entanglement advantage persisted in the presence of depolarising noise in both the channels and entangling gates provided the noise was sufficiently small. However, this work considered the rather unrealistic case of perfect state transmission, i.e., for lossless channels.

\begin{figure}[htb]
\includegraphics[width=0.25\textwidth]{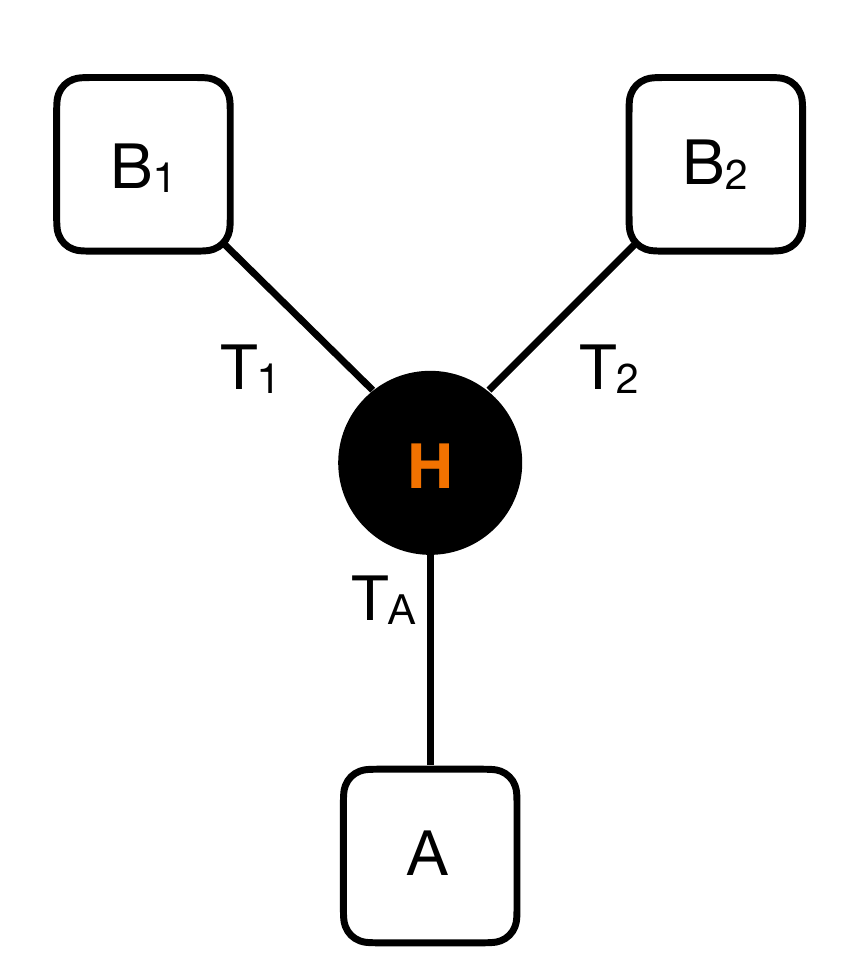}
\caption{A tri-partite quantum communication scenario between a dealer, here player $A$, and two other players, $B_1$ and $B_2$. This quantum network with a central router $H$, which is able to produce and entangle qumodes, exemplifies a network with a bottleneck. The CV graph state used in the multi-partite entanglement QSS protocol, can be distributed in a single network use (i.e., each channel transmits a single qumode only), while $(n-1)$ uses of the network are necessary in the bQSS protocol. \label{bottleneck}}
\end{figure}

Here, we will consider CV QSS over a bottleneck network for the simplest non-trivial scenario with $n=2$ Bobs (Fig.~\ref{bottleneck}). To map out the ultimate limits to performance advantage in this scenario we first present asymptotic key rates using finite squeezing for the graph states but with all other parameters being ideal. The links are modelled as pure loss channels which are an excellent first approximation to a fibre optic network. 

There are several considerations that are specific to the fact that this is CV protocol. The first is that there are two different ways for the entangled state to be distributed across the bottleneck network, although both only require a single network use. The hub can simply create an entangled state and send one mode to each player (Hub-Out) or, alternatively, one player could create a two-mode graph state and send one half to $H$ where it will be entangled with a third mode and then distributed amongst the remaining players (player-in). For a DV protocol, a pure loss channel only effects the probability of photon arrival and the two methods would yield identical states with the same transmission probabilities. However, for the CV case the entangling gates do not commute with the lossy channels and a different entangled state is distributed depending upon which network coding method is employed. Note that the method where the initial entangled state is created at the central hub substantially more practical as it can be achieved using only offline squeezing. In other words, the required entangled state can be made beginning with three squeezed vacuum modes that are passed through an appropriate linear optical unitary.

The second CV specific point is that, for similar reasons, with a player-in strategy it matters whether the player who initially transmits the state is the dealer (who is the reference player in the sense it is their measurements that will make up the secret key) or one of the Bob's. This is essentially the same asymmetry observed in CVQKD where one finds different keyrates for so-called direct and reverse reconciliation \cite{Grosshans:2002p5549}. A similar effect occurs even with a Hub-Out strategy if the hub prepares an asymmetric graph state (e.g., a line graph as opposed to a fully symmetric graph). Third, for a fixed, finite amount of available squeezing there is in fact a whole family of CV graph states where the squeezing is divided between the initial squeezed vacuum states and the $\mathrm{CPHASE}$ gates that create the graph state. This should be optimised over for a given secret sharing protocol. Lastly, CV graph states generally have asymmetric quadrature correlations and therefore it is crucial to make an optimal choice for which quadrature is encoded with the key and which is used as the check. This optimal choice is dictated by the correlation structure of the graph state (see Appendix~\ref{asympapp} for a detailed explanation)

Since offline squeezing is much more practical with present technology, here we will only consider a Hub-Out strategy, where a three-mode line graph is created to implement a $(2,2)$-threshold scheme over a bottleneck network of lossy fibre-optical channels. For the squeezing resource we will assume an initial available squeezing of 15dB corresponding to the state of the art values for measured vacuum squeezing \cite{Vahlbruch:2016jf}. The term measured squeezing refers to the fact that the actual squeezing generated by state of the art nonlinear processes is typically much higher ($>$20dB) but due to system losses the real output is a slightly mixed state which produces a smaller measured squeezing. We model this setup in detail later, but for now, 
we approximate the output as a pure state but with a degree of squeezing limited to the measured value. Given a maximum initial, offline squeezing value, one can use the Bloch-Messiah decomposition to construct a family of approximate graph states where the is a freedom to divide this squeezing `budget' between the entangling gates and the initial squeezed vacuum states that would appear in the equivalent canonical construction. For all key rates plotted here we will optimise over this choice (see Appendix~\ref{asympapp} for details).

\begin{figure}[htb]
\includegraphics[width=0.45\textwidth]{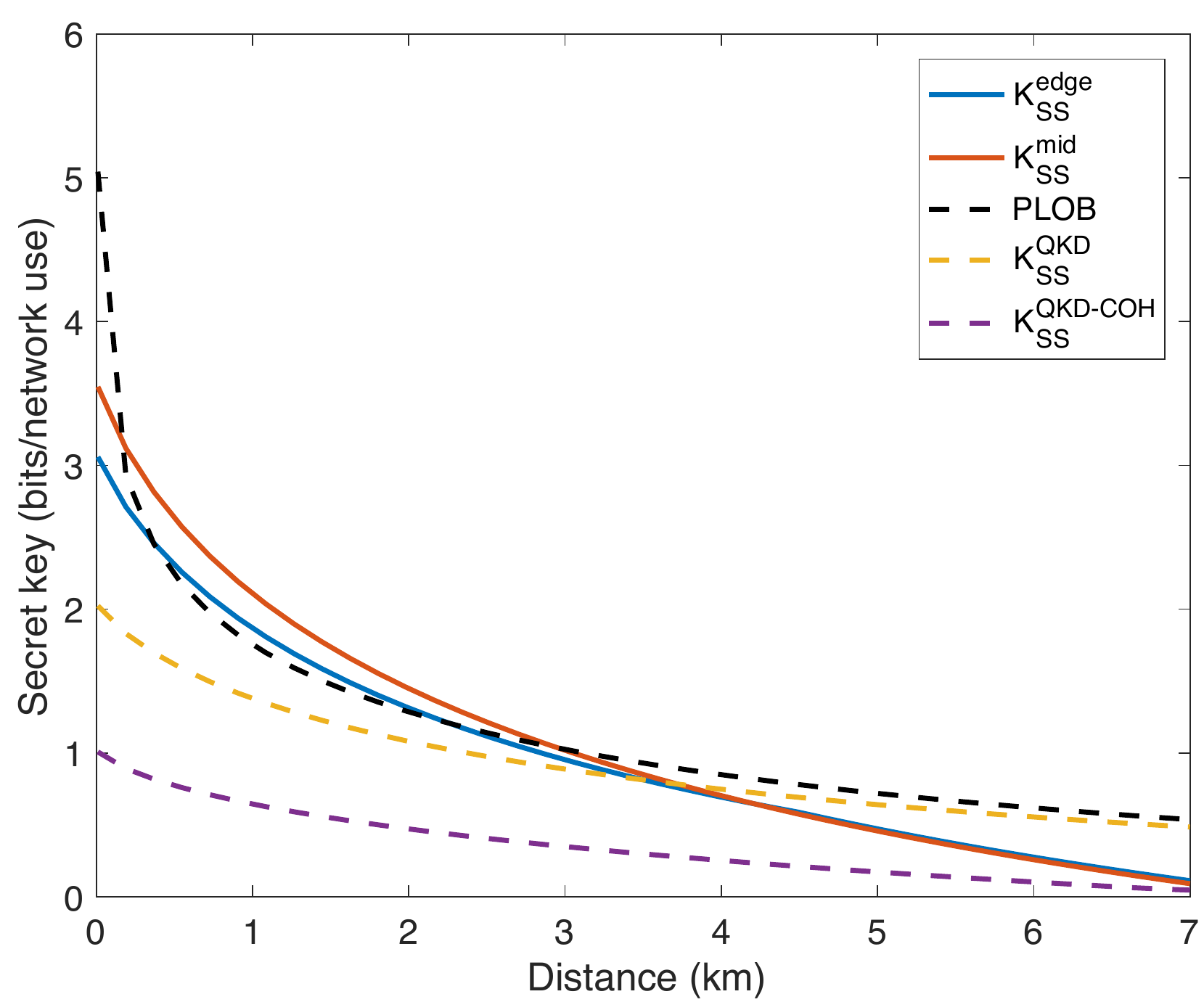}
\caption{Comparison of secret sharing rates as a function of transmission radius for a symmetric lossy bottleneck network between entanglement based protocols (solid lines), the corresponding CVQKD based protocol based on squeezed states (dashed yellow) coherent states (dashed purple) or the best possible bi-partite point-to-point protocol (dashed black). The maximum available squeezing is 15dB and the graph state generation process is optimised with respect to this limit. \label{asympcomp}}
\end{figure}

For simplicity, we will consider a symmetric network with the players situated at an identical distance from the central hub such that $T_A=T_1=T_2=T$. In Fig.~\ref{asympcomp}, we plot 
the secret sharing rate given by Eq.\ (\ref{kasymp}) as a function of the distance in kilometres, $d$, which is related to the transmission via $T = 10^{-.02 d}$.  A multi-partite entangled strategy also enjoys a qualitative advantage over any QKD based implementation in that the dealer, can in fact be chosen after the quantum states have been distributed. However, the choice of dealer will effect the performance. For a three mode line graph there are two possible configurations depending on whether the dealer possesses the middle mode or one of the edge modes (due to the symmetry of our network the two edge-modes are result in identical rates). Interestingly, we see that for all transmissions it is favourable for the dealer to be sent the middle node in the chain.

Turning to the comparison with bQSS, we can straightforwardly compute a benchmark \cite{BenchmarkingReview} by evaluating the secret sharing rate for a scheme based upon a bi-partite 
CVQKD protocol between the dealer and each player with the same squeezing resources, and then dividing by the number of additional network uses required. The asymptotic rate of an $(n,n)$-scheme over the same symmetric bottleneck network is then
\eqn{K_{\mathrm{bSS}} = \min_{B_i \in \mathcal{B}} \frac{1}{n} \bk{I(X_A:X_{B_i}) - \chi(X_A:E\mathcal{B}/B_i )}.\label{bss}} 
Note that we can compute the key rate for more than one Gaussian CVQKD protocol. The two natural choices are: (i) the optimal protocol where Alice sends Gaussian modulated squeezed states (or equivalently homodynes one half of a two-mode squeezed vacuum) and the Bobs homodyne detect; the protocol where Alice sends Gaussian modulated coherent states and the Bob's homodyne detect. This latter protocol is less loss tolerant, but requires no squeezing and is thus very cheap, robust and often favoured in field implementations so we also include it as a comparison. All key rates are computed in Appendix \ref{asympapp}.

In Fig.~\ref{asympcomp}, it can be seen that graph state secret sharing achieves a higher secure rate for short distances, outperforming the corresponding squeezed state bQSS protocol up until a transmission radius of over 3 km and the coherent state protocol till 7 km (meaning the parties could be as far away as 7 and 14 km respectively). For an even more dramatic illustration of the potential benefits of multi-partite entanglement, 
\begin{equation}
K_{\mathrm{PLOB}} = -1/2\log_2\bk{1-T^2}. \label{PLOB}
\end{equation}
This represents the maximum possible rate for any QKD based secret 
sharing protocol over the same network, even including unlimited squeezing or input energy. Remarkably, an entanglement based protocol with finite squeezing can outperform even this benchmark for a transmission radius of up to approximately 2.5km.

The relative performance of the multi-partite protocol,  being superior for low environmental degradation but inferior for higher transmission losses, is consistent with previous work \cite{Epping:2017dx} and 
can  be  understood  as  follows.   For  bQSS  schemes,  there  is  only  ever  one  channel  in  use  for a single QKD protocol, which is then leveraged into the full QSS protocol.  In the multi-partite, the malicious parties can collect information from all channels simultaneously, which leads to much worse performance as the loss of the individual network links grows higher. This is why the multi-partite advantage vanishes when the loss is above a certain threshold.

\begin{figure}[htb]
\includegraphics[width=0.45\textwidth]{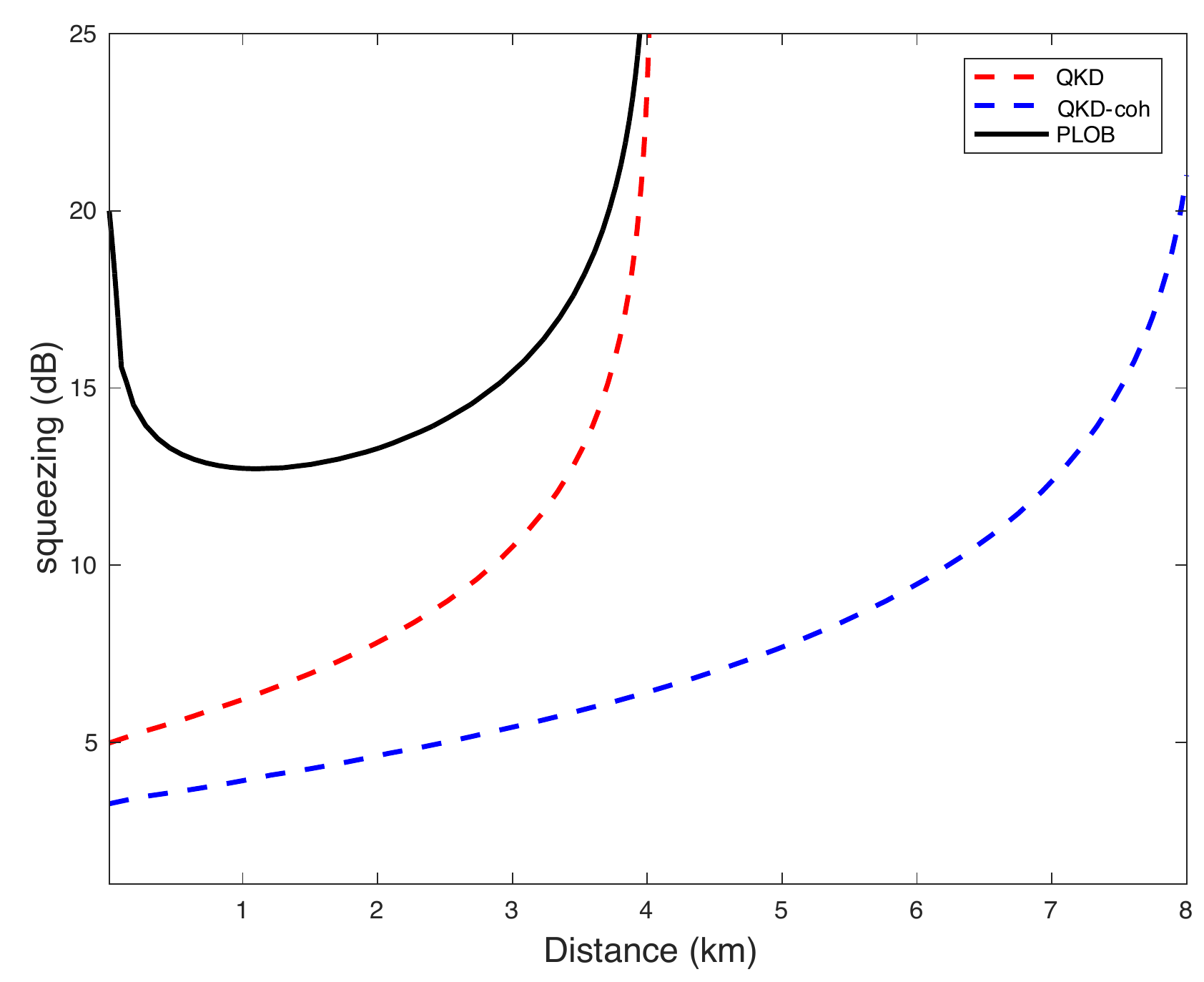}
\caption{Asymptotic advantage region for entanglement based secret sharing. For the network given in Fig.~\ref{bottleneck} our entanglement based protocol outperforms an optimal, infinite energy bi-partite QKD based protocol for all values of initial squeezing and transmission radius above the solid black line. The advantage region compared to a squeezed (dashed red) and coherent (dashed blue) state CVQKD protocol with the same energy/squeezing resources is also shown. \label{advantangeregion}}
\end{figure}

We further investigate the parameter regimes where our multi-partite strategy enjoys and advantage over the various bi-partite benchmarks in Fig.~\ref{advantangeregion} by mapping out the contours of squeezing and transmission radius for which the key rates coincide. Firstly, these curves show that for this bottleneck scenario, QSS protocols are only superior for intra-city networks with a radii of 3-6km. On the other hand, whilst beating the ultimate PLOB limit requires at least 12.5 dB of squeezing or more, values of around 6dB are sufficient to to surpass the comparable CVQKD protocols based on squeezed or coherent states up to radii of 2 km and 4 km respectively.

Crucially, protocols utilising Gaussian CV entanglement are deterministic and run at the same raw clocks speed as standard CVQKD methods. This means that these advantages will directly hold in terms of secret bits per second. This is in contrast with most optical DV GHZ experiments where the non-deterministic nature of the state creation process means that generation rates fall substantially as the number of parties grows beyond the bi-partite case. This means that for current DV implementations that an advantage "per channel use" will not necessarily manifest as an advantage "per time". Motivated by this potential for a real advantage with current CV systems we now analyse an implementation with realistic, present-day devices in a composable, finite-size setting.

Firstly, we will consider imperfect reconciliation efficiency such the amount of information leakage becomes $\ell_{\mathrm{IR}} = H(X_A) - \beta I(X_A:X_B)$ where $\beta\leq 1$ quantifies the fraction ideal Shannon-limited mutual information achieved by a given error-correction code. Secondly, for homodyne protocols the only known composable security proofs rely on entropic uncertainty relations that, as mentioned before, are provably not tight. Thirdly, real fiber optic channels are not exactly pure loss channels, and instead exhibit a small amount of excess thermal noise. Finally, as well as accounting for the finite dynamic range and detector resolution we also model realistic imperfections in the state generation including cavity escape losses, finite detector efficiency and losses coupling into the transmission fiber. All values are taken from reported experimental demonstrations and a full description of the model can be found in Appendix~\ref{expmodel}.

\begin{figure}[htb]
\includegraphics[width=0.45\textwidth]{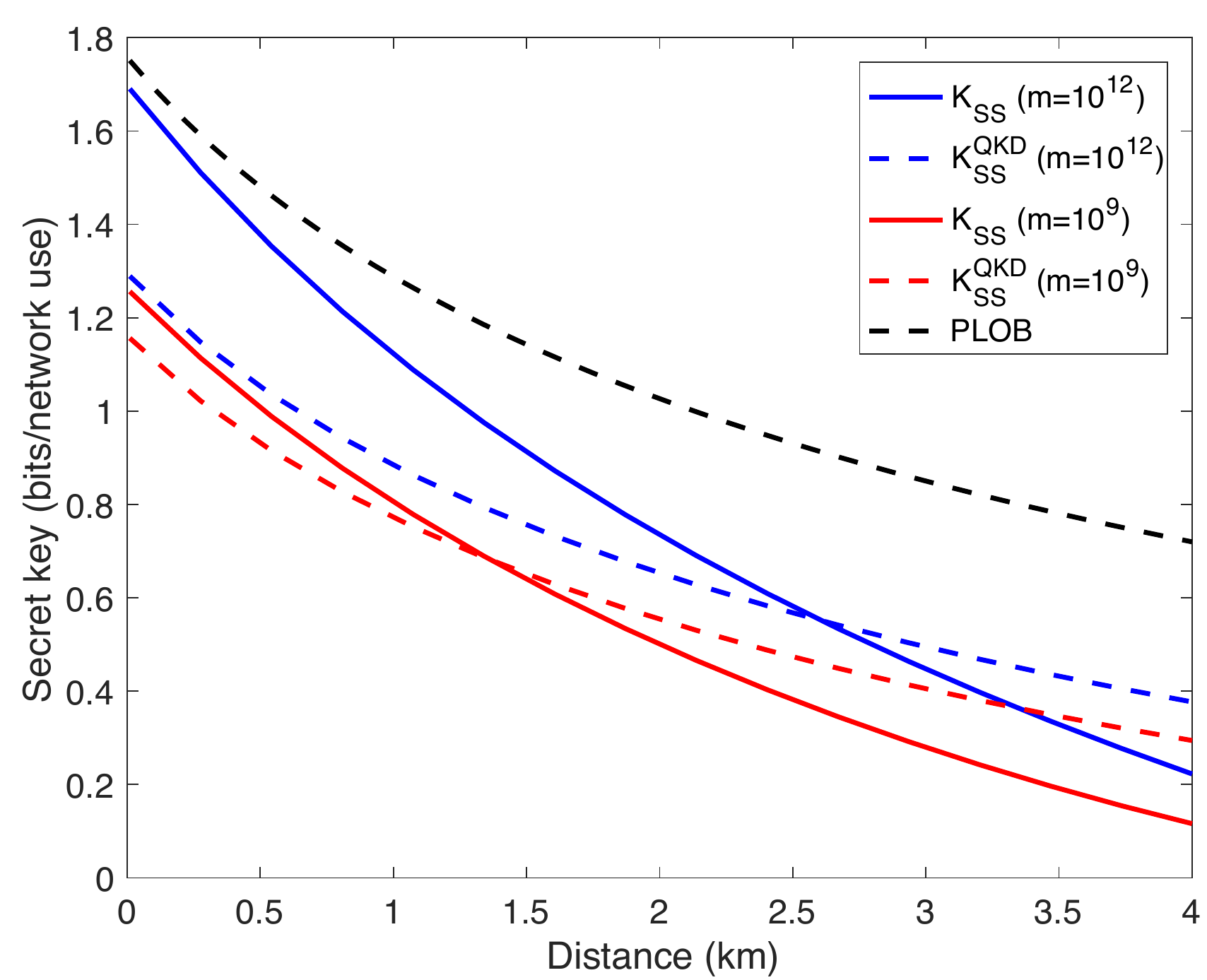}
\caption{Comparison of ($\epsilon_s + \epsilon_c$)-secure secret sharing rates as a function of transmission radius for a symmetric bottleneck network of thermal loss fibre channels between entanglement based protocols (solid lines), the corresponding CVQKD based protocol based on squeezed states (dashed lines) for block sizes of $10^{12}$ (blue) and $10^9$ (red). Other parameters: $\epsilon_s = \epsilon_c = 10^{-9}$; reconciliation efficiency $\beta = 0.98$ \cite{Pacher:2016te}; escape efficiency $\eta_{\mathrm{es}} = 0.99$, detector efficiency $\eta_d = 0.99$, inferred pure squeezing 23.3 dB \cite{Vahlbruch:2016jf}; fibre coupling efficiency 0.95 \cite{Gehring:2015ie}; transmission $T = 10^{-0.02 d}$, excess noise $\xi = 0.002$ \cite{Jouguet:2013p8197}, detector resolution $\delta_\mathbb{X} = 0.1$, $\delta_\mathbb{P} = 0.4$ and range $M_{\mathbb{P}/\mathbb{X}}=25$ \cite{Gehring:2015ie}, energy test parameters, $T_e = 0.99$, $\alpha = 28$; security proof constants $\epsilon_1 = 4\times 10^{-11}$, $\epsilon_2 = 4\times 10^{-20}$, the probability of a key generation round, $p$, is numerically optimised. The PLOB bound of a optimal bi-partite protocol including transmission and experimental losses but with otherwise perfect equipment is also plotted (dashed black). \label{finitecomp}}
\end{figure}

To make a fair comparison with a bQSS protocol, we also compute the composable finite-size CVQKD key rate for an implementation with the same level of available squeezing and experimental imperfections. It is important to emphasise that, even when fairly allocating resources in this way, it is not immediately obvious that the multi-partite advantage will survive. A CVQKD protocol can be made more efficient  (it is possible to avoid losing rounds due to basis mismatch via pre-shared key) and even with identical noise levels for squeezers, fibre-couplers etc, a QKD based implementation uses less devices in total and hence introduces less noise. A detailed explanation of the comparisons and calculation of the QKD-based secret fraction is given in Appendix~\ref{finres}. Lastly, although it is arguably unfair to compare these finite-sized results to the asymptotic PLOB bound, we nevertheless include it as an instructive upper bound the best performance possible for bQSS. As well as infinite communication rounds, the standard PLOB bound holds in the limit of perfect devices. To make a fairer comparison and more accurately highlight the advantage of multi-partite entanglement, we make one modification towards realism in the PLOB bound by setting the loss equal to the total effective loss in the realistic implementation. In other words, in Fig.~\ref{finitecomp} we have assumed that the losses from fibre coupling, squeezing cavity and detectors are unavoidable and the PLOB bound is evaluated via Eq.~(\ref{PLOB}) but with a transmission of $\eta_f(T\eta_d\eta_s)^2$ instead of $T^2$. This still corresponds to a protocol with perfect reconciliation efficiency, detector range and resolution, an absence of any excess noise and infinite encoding energy and so can be taken to be an optimistic upper bound for the performance of a bQSS scheme.

In Fig.~\ref{finitecomp}, we find that a realistic, finite-size, multi-partite secret sharing scheme can no longer surpass the PLOB bound. This is perhaps unsuprising as the PLOB bound is an inherently asymptotic result. The loss of performance due to finite-size effects in our secret sharing protocol in comparison to standard QKD is discussed in more detail in Appendix C.
However, when making the more reasonable comparison to the equivalent realistic bQSS protocol, we see that for sufficiently large block sizes of $m=10^{12}$ there is a quantitative advantage for a transmission radius of up to 2.5~km. A lesser advantage persists for shorter block sizes, but we see that for $m=10^9$ the advantage is much smaller and the region is only up to around 1.5~km.

The fact that we first reduced the security of our multi-partite protocol to a minimisation over bipartite protocols is critical here. The Gaussian extremality results have only been proven to hold in a bipartite setting, so it has been crucial that we first made this reduction and in order to apply them.


\section{Conclusions and outlook}

In this work, we have provided a security proof for an important multi-partite quantum communication scheme: A $(n,k)$-threshold scheme for sharing classical secrets with multi-partite entanglement in the composable, finite-sized setting. This protocol is secure against general quantum attacks, including participant attacks. When applied to the original secret sharing scheme the proof never certifies a positive key rate but a CV scheme based on Gaussian graph states shows robust performance. Moreover, we showed that for the specific example of a $(2,2)$ scheme implemented over a three-party bottleneck fibre-optic network, the multi-partite scheme exhibits superior performance for intra-city transmission distances.

In the limit of a large number of communication rounds, 
this scheme outperforms not only a bi-partite protocol based upon a CVQKD protocol with the same squeezing resources, but even surpasses 
implementation-agnostic 
and overly optimistic 
bounds. Indeed, it even outperforms the PLOB bound which represents the ultimate limit for any point-to-point private communication. Perhaps most importantly, we show that an advantage persists even in the finite-sized regime for a implementation modelled on existing, state-of-the-art squeezing experiments. It is worth noting that in the advantage regime the key rates are also always greater than 1 bit per channel use, therefore automatically also outperforming recent advances in so-called twin-field QKD \cite{Lucamarini:2018if} which can also break the PLOB bound. A demonstration of this proposal, which should be possible with present day technology, would represent a watershed demonstration of a quantitative advantage for multi-partite entanglement based quantum communication using realistic channels.

There are several avenues for future research opened up by this work. Perhaps the most pressing open question is a thorough investigation of how this scheme scales to larger numbers of players over more complicated network topologies (e.g., a butterfly network). Also, for reasons of practicality, in this work we focused only on implementations using offline squeezing but preliminary results suggest that performance could be improved in inline squeezing resources were to become readily available. There is also an in-principle qualitative advantage to entanglement based secret sharing, which is that the identity of the dealer can be chosen after state distribution, albeit at the price of reduced performance. A further interesting direction is the extent to which other quantum coding techniques such as local complementation \cite{HahnPappaEisert} can be used to ameliorate this problem and fully exploit this added flexibility \cite{CVLCprep}. 

Whilst our results indicate implementations with near term technology will only be feasible over metropolitan distances, in future, sophisticated quantum networks \cite{WehnerQuantumInternet} that include repeater stations \cite{Briegel:1998jd,Duan:2001ke,Dias:2017jk,Furrer:2018im} or ones building on fault-tolerant protocols \cite{Munro:2012gu,Azuma:2015vq} may render a multi-partite advantage achievable over much longer distances.

Although this proof fails to give positive key rates when applied to the original HBB proposal with GHZ states, variants of this scheme could still demonstrate useful performance via our proof method \cite{GHZprep}. On a broader perspective, it is the hope that this
work stimulates further studies of protocols
making use of multi-partite entangled resources that achieve a genuine advantage over point-to-point protocols, providing further perspective to the field
of quantum communication beyond point-to-point schemes.

\begin{acknowledgements}
The authors thank M.~J.\ Hoban, J.\ Memmen, 
and H.~M.\ Chrzanowski for helpful discussions and careful proofreading. N.~W.\ acknowledges funding from the European Unions Horizon 2020 research and innovation programme under the Marie Sklodowska-Curie Grant  Agreement No.750905. Both N.~W.\
and J.~E.\ thank the Q.Link.X
and QR.X from the BMBF in Germany and the DFG priority program “Compressed Sensing in Information Processing - Phase 2 (CoSIP2)” for support.   \end{acknowledgements}

\begin{appendix}
\section{Secret sharing rates with CV graph states \label{CVgraphs}}

In this section, we will
review the formalism of bosonic Gaussian states as it is needed to describe the protocols considered here and also detail the noise models used.
\subsection{Preliminaries}
Although Gaussian states are supported on infinite dimensional Hilbert spaces, they can be completely described by a finite number of parameters, namely their first and second moments. Similarly, Gaussian operations can be compactly captured by symplectic transformations. For a detailed discussion the reader should consult Refs.\  \cite{Weedbrook:2012p5160,Continuous}.

A bosonic system can be described in terms of appropriate \emph{creation and annihilation operators}. For an $N$ mode system it can be convenient to group these into vectors
\begin{eqnarray}
    \hat{\mathbf{a}}&:=&\left(\hat{a}_{1}, \ldots, \hat{a}_{N}\right)^{\intercal},
\end{eqnarray}
with the creation operators being the Hermitian conjugates of these operators.
Such systems can equivalently represented by the \emph{quadrature operators} defined by $\hat{a}_{k}:=\frac{1}{2}\left(\hat{x}_{k}+i \hat{p}_{k}\right)$ 
for $k=1,\dots, N$, or equivalently
\eqn{\hat{x}_{k}:=\hat{a}_{k}+\hat{a}_{k}^{\dagger}, \quad \hat{p}_{k}:=i(\hat{a}_{k}^{\dagger}-\hat{a}_{k})}
which for an $N$-mode system we can write as
\eqn{\hat{\mathbf{r}}:=(\hat{x}_{1}, \hat{p}_{1}, \ldots, \hat{x}_{N}, \hat{p}_{N})^{\intercal}.\label{x1p1}}
Note that by choosing these particular pre-factors linking the quadrature operators to the annihilation and creation operators we are setting $\hbar = 2$, which corresponds to $\left[\hat{x}, \hat{p}\right]=2 i$ 
and will ensure that the vacuum variance is normalised to 1. The symplectic form associated with the ordering defined by (\ref{x1p1} is
\eqn{\boldsymbol{\Omega}=\bigoplus_{k=1}^{N} \boldsymbol{\sigma}, \quad \boldsymbol{\sigma}=\left(\begin{array}{cc}
0 & 1 \\
-1 & 0
\end{array}\right).}
One can also use a different operator ordering convention and define a vector of quadrature operators,
\eqn{\hat{\mathbf{q}} := \bk{\begin{array}{c}
    \hat{\mathbf{x}}  \\
     \hat{\mathbf{p}} 
\end{array}}, \label{x1x2} }
where
\eqn{\hat{\mathbf{x}} =(\hat{x}_1,\dots,\hat{x}_N)^\intercal,\hs \hat{\mathbf{p}} = (\hat{p}_1,\dots ,\hat{p}_N)^\intercal.}
The \emph{symplectic form} reflecting the canonical commutation relations takes in this convention the form
\begin{equation}
    \omega=2 
    \mat{0}{-\mathbf{I}_N}{\mathbf{I}_N}{0}.
\end{equation}
The two conventions are naturally related by an appropriate permutation operation. For most of this work we will use the convention in (\ref{x1p1}) however sometimes it can be more convenient to adopt (\ref{x1x2}) and it will be made clear when this is done.

An arbitrary $N$-mode Gaussian state $\hat{\rho}$ can be 
completely specified by a 
\emph{vector of first moments},
\eqn{\overline{\mathbf{r}}:=\langle\hat{\mathbf{r}}\rangle=\operatorname{Tr}(\hat{\mathbf{r}} \hat{\rho}),}
the displacements in phase space,
and a \emph{covariance matrix (CM)} 
that captures the second moments. This covariance matrix $\mathbf{\Gamma}$ has entries
\eqn{\Gamma_{i,j}:=\frac{1}{2}\left\langle\left\{\Delta \hat{r}_{i}, \Delta \hat{r}_{j}\right\}\right\rangle.}
Covariance matrices of
multi-partite systems, which we will label with subscripts, can be written a convenient block form. For example, an arbitrary tri-partite system of a state $\hat{\rho}_{A,B,C}$ can be 
written as
\eqn{\mathbf{\Gamma}_{A,B,C} = \matthree{\mathbf{\Gamma}_A&\mathbf{C}_{A,B}&\mathbf{C}_{A,C}}{\mathbf{C}_{A,B}^\intercal &\mathbf{\Gamma}_B  &\mathbf{C}_{B,C}}{\mathbf{C}_{A,C}^\intercal&\mathbf{C}_{B,C}^\intercal&\mathbf{\Gamma}_C}.}
Tracing out a subsystem simply corresponds to discarding the appropriate part of the total CM
and considering a principle sub-matrix, so that, for example, the CM of the reduced state $\hat{\rho}_{A,C} = \mathrm{tr}_B(\hat{\rho}_{A,B,C})$ is given by
\eqn{\mathbf{\Gamma}_{A,C} =\mat{\mathbf{\Gamma}_A}{\mathbf{C}_{A,C}}{\mathbf{C}_{A,C}^\intercal}{\mathbf{\Gamma}_C}.}
Measuring out a quantum subsystem via a homodyne detection is given by the appropriate Schur complement
\cite{Fiurasek:2002p467,Eisert:2002p466,Giedke:2002p468}. In the above situation if, instead of being traced out, the mode $B$ is measured in the $\hat{x}$ quadrature, 
the conditional CM is given by
\eqn{\mathbf{\Gamma}_{A,C|x_B} = \mathbf{\Gamma}_{A,C} - \mathbf{C}\bk{\mathbf{X\Gamma}_B\mathbf{X}}^\mathrm{MP}\mathbf{C}^\intercal\label{schur}}
where $\mathrm{MP}$ denotes the Moore-Penrose matrix inverse,
\eqn{\mathbf{C}^\intercal = \bk{\mathbf{C}_{A,B} \mathbf{C}_{B,C}}\nn}
is the total correlation matrix between $B$ and the joint $A,C$ system and $\mathbf{X} =
\text{diag}(1,0)$
(for a $\hat{p}$ measurement we would instead use $\mathbf{P} =
\text{diag}(0,1)$). The conditional first moment is given by
\eqn{\mathbf{r}_{A,C|x_B} = \mathbf{r}_{A,C} + \mathbf{C}\bk{\mathbf{X\Gamma}_B\mathbf{X}}^\mathrm{MP}(\mathbf{m}-\mathbf{r}_B)\label{rcond}}
where $\mathbf{m} = \text{diag}(x_B,0)$ is the measurement vector where the non-zero entries are Bob's measurement outcomes (in this case in the $\hat{x}$ quadrature). The analogous result holds for conditioning on a $\hat{p}$ measurement.

An arbitrary Gaussian unitary can be compactly represented by
matrix from the
\emph{real symplectic group}
$\mathbf{S} \in Sp(2N,\mathbb{R})$ 
so a real matrix satisfying
\begin{equation}
    \mathbf{S\Omega
    S^\intercal =
    \Omega},
\end{equation}
and a vector $\mathbf{d} \in \mathbb{R}^{2N}$ that together define a corresponding affine transformations of the first moments and a symplectic transformation of the CM given by
\eqn{\overline{\mathbf{r}} \mapsto \mathbf{S}\overline{\mathbf{r}} + \mathbf{d}, \hs \mathbf{\Gamma} \mapsto \mathbf{S\Gamma S^T}.}
The specific Gaussian operations we will require for our calculations are \emph{single mode squeezing operations} in the $\hat{x}$ quadrature with squeezing parameter $r>0$
\eqn{\mathbf{S}(r) = \mat{e^{-r}}{0}{0}{e^r}}
and a \emph{beam-splitter with transmissivity}
$T\in [0,1]$
\eqn{\mathbf{BS}(T) =\left(\begin{array}{cc}
\sqrt{T} \mathbf{I}_2 & \sqrt{1-T} \mathbf{I}_2 \\
-\sqrt{1-T} \mathbf{I}_2 & \sqrt{T} \mathbf{I}_2
\end{array}\right)}
where $\mathbf{I}_2$ is the $2\times 2$ identity matrix. Finally, we require a two mode entangling gate sometimes called a $\mathrm{CPHASE}$ gate or a $\mathrm{CZ}$ gate by analogy with qubit systems. A $\mathrm{CZ}$ gate gate of strength $g$ is described by the symplectic matrix
\eqn{\mathbf{CZ}(g) = \left(
\begin{array}{cccc}
 1 & 0 & 0 & 0 \\
 0 & 1 & g & 0 \\
 0 & 0 & 1 & 0 \\
 g & 0 & 0 & 1 \\
\end{array}
\right).\label{CZgate}}

For one and two mode operations acting on larger systems multi-mode systems we will use subscripts to denote the target modes, and the necessary padding with identity matrices defined implicitly as appropriate, e.g., a single mode squeezing on mode $A$ of a joint $A,B$ system would give rise to
\eqn{\mathbf{S}_A(r) =\mat{\mathbf{S}_A(r)}{\mathbf{0}}{\mathbf{0}}{\mathbf{I}_2}= \matfour{e^{-r}&0&0&0}{0&e^r&0&0}{0&0&1&0}{0&0&0&1}}
where $\mathbf{0}$ is a $2\times 2$ matrix of zeroes. Similarly, a beam-splitter operation between modes $A$ and $B$ of a three mode system would be written as
\eqn{\mathbf{BS}_{A,B}(T) = \mat{\mathbf{BS}(T)}{\mathbf{0}}{\mathbf{0}}{\mathbf{I}_2}.}

\subsection{CV graph states} 

Equipped with this framework, we can state the definition of \emph{CV graph states} \cite{PhysRevA.76.032321,PhysRevLett.97.110501,PhysRevA.79.062318} as continuous analogues of \emph{graph states} \cite{Hein04,PhysRevA.68.022312} as instances of stabilizer states. At the heart of the 
concept of a CV graph state is an adjacency matrix
\begin{equation}
\mathbf{A = A^\intercal},
\end{equation}
of a weighted graph having zero entries for
pairs of modes that are not connected and a positive value for pairs modes that are connected. By convention, each mode is initialised in $p$-squeezed vacuum state and the adjacency matrix of a
weighted graph captures the interaction 
pattern. The role of the adjacency matrix in state generation is most apparent if we switch to the ordering convention of Eq.~(\ref{x1x2}).
The symplectic transformation implementing an imperfect CV graph state in this convention is
\eqn{ 
\mathbf{\tilde{G}}=\mat{\mathbf{1}_N}{\mathbf{1}_N}{\mathbf{A}}{\mathbf{1}_N}
\mat{e^{-r} \mathbf{1}_N}{0}{0}{e^{r}\mathbf{1}_N},\label{Sgraph}}
where $r$ parameterises the initial squeezing and the tilde is to emphasise that this matrix is written in a different ordering convention. It takes a moment of thought that 
these matrices satisfy
$S\in Sp(2N,\mathbb{R})$.
These are imperfect CV graph states
\cite{PhysRevA.79.062318,Ohliger:2010kv}, and become infinite energy improper quantum states in the limit $r\rightarrow\infty$.
Such imperfect CV graph states are at the heart of our formalism. Another useful way to conceptualise CV graph states are via their \emph{nullifiers} which are collection of $N$ multi-mode observables defined uniquely for a given adjacency matrix by the equations,
\eqn{\mathbf{\hat{n}} = \mathbf{\hat{p}} - \mathbf{A\hat{x}}\label{nullifiers}}
where we are again using the definitions in Eq.~(\ref{x1x2}). One way to understand the correlation structure of these graph states is to think that the original squeezing is now distributed in a nonlocal observable made up of quadratures from the various nodes of the graph state. The perfect graph state arising from infinite squeezing therefore results in maximum correlation and it is straightforward to show that \cite{PhysRevLett.97.110501,PhysRevA.79.062318}
\eqn{\lim_{r\rightarrow \infty}  \mathbf{n} = \mathbf{0}}
This will become useful later when choosing the optimal secret sharing strategy.

\begin{figure}[htb]
\includegraphics[width=0.35\textwidth]{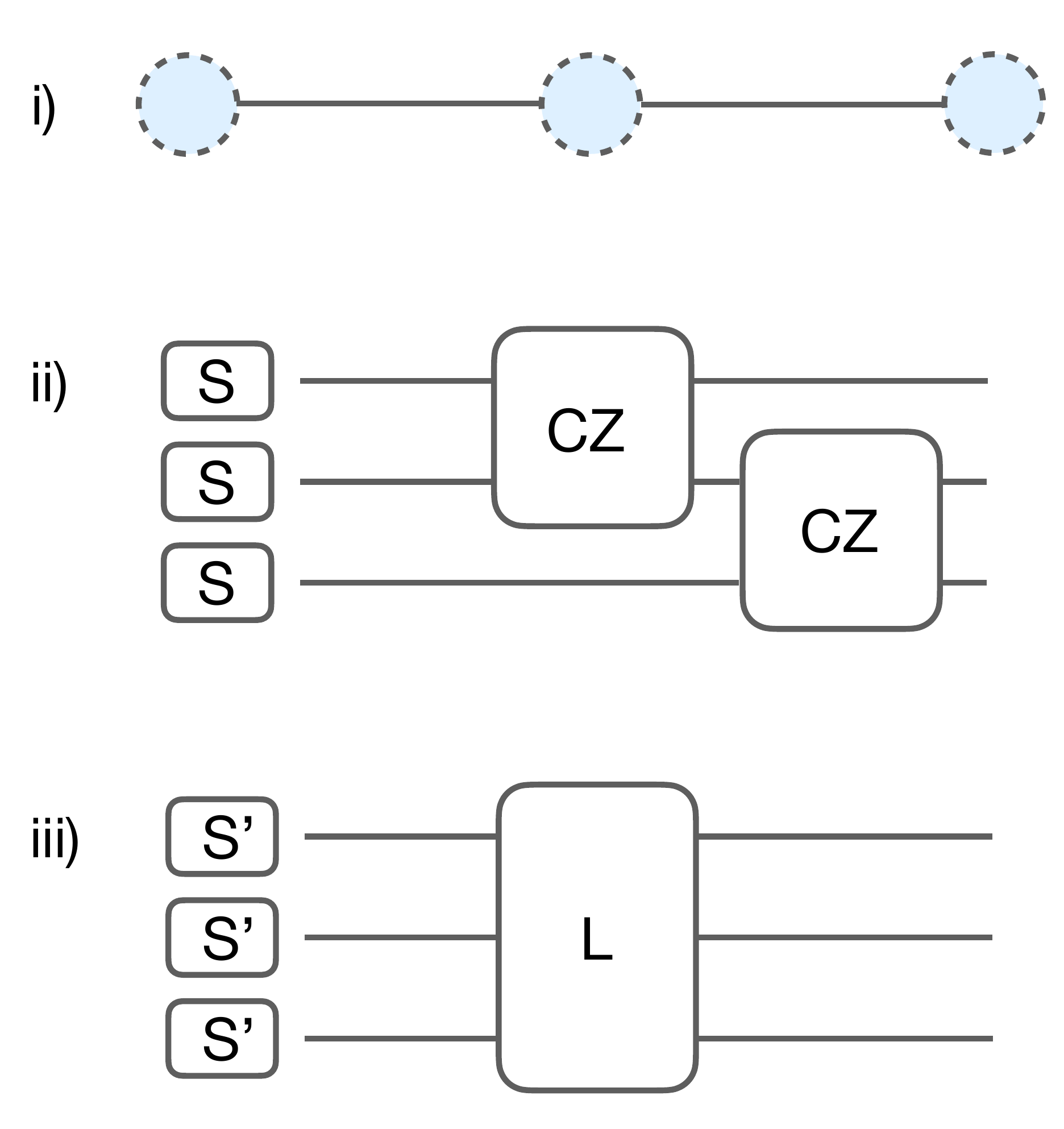}
\caption{i) Graphical representation of a tri-partite line graph. ii) Canonical generation method, where each graph vertex is initialised in a squeezed vacuum state and the each edge is created via a CZ gate. iii) Practical generation via offline squeezing. Here the squeezers, $S'$, will generally be stronger than those appearing in the canonical construction in ii).  \label{CV_graph}}
\end{figure}

The canonical method to realise a given graph state (Fig.~\ref{CV_graph}.(ii)) is to implement a CZ gate for each edge in the graph. The weight of each edge corresponds to the strength, $g$, of the entangling gate as per Eq.\ (\ref{CZgate}). Note that a perfect graph state emerges by taking the infinite squeezing limit in the initial squeezed vacuum states as distinct from the taking the limit of infinite weight of the entangling gates. Taking the limit $g\rightarrow \infty$ in the CZ gates would \emph{not} correspond to a perfect graph state. The tri-partite line graph we will be using can therefore be written
\eqn{\mathbf{G}_{L} &=& \mathbf{CZ}_{B,C}(g)\cdot \mathbf{CZ}_{A,B}(g)\cdot\mathbf{S}_C(-r) \cdot\mathbf{S}_B(-r) \nn\\
&\times &\mathbf{S}_A(-r). \label{canonical}}
Finally, a more practical construction is to prepare the graph state via offline squeezing \cite{PhysRevA.76.032321}. This is done via the \emph{Bloch-Messiah decomposition} which allows an arbitrary Gaussian unitary to be decomposed into a passive, linear-optical interferometer followed by a single-mode squeezing operations and a second passive interferometer \cite{Braunstein:2005p6353,Cariolaro:2016du}. When starting from vacuum state, as we are here, the first interferometer can be ignored and an arbitrary graph state can be prepared as per Fig.~\ref{CV_graph}.(iii) by a layer of single-mode squeezers and a final passive unitary. The squeezers in the Bloch-Messiah composition will necessarily be stronger than the initial squeezers in the canonical construction since they must also incorporate the squeezing that would go into generating the CZ operations. Following 
Ref.~\cite{Cariolaro:2016du}, 
we obtain the following decomposition for the graph state given by (\ref{canonical}),
\eqn{\mathbf{G}_{\mathrm{Bloch}} = \mathbf{L}\cdot \mathbf{S}_A(-r_A)\cdot\mathbf{S}_B(-r_B)\cdot\mathbf{S}_A(-r_B) \label{Gbloch}}
where $r_A = r$, 
\eqn{r_B \ee r_C = \log \left[\frac{1}{2} \left(\sqrt{\left(2 g^2+1\right) e^{2 r}+e^{-2 r}-2} \right. \right. \nn\\ 
&+&\left. \left.\sqrt{\left(2 g^2+1\right) e^{2 r}+e^{-2 r}+2}\right)\right] \label{rbloch}}
and $\mathbf{L}$ is the symplectic transform of the passive interferometer. This can be obtained by essentially carrying out a series of singular value and eigenvalue decompositions of the symplectic transform describing the canonical generation of the target graph state. These can be readily obtained via a mathematical software package and it can also vbe checked that these procedures satisfy $\mathbf{G}_{\mathrm{Bloch}}\cdot \mathbf{G}_{\mathrm{Bloch}}^\intercal = \mathbf{G}_{L} \cdot \mathbf{G}_{L}^\intercal $ as required. To further simplify experimental implementation this linear optical unitary can further be simplified into a network of beamsplitters and phase shifters via the Reck \cite{Reck:1994dz} or Clements \cite{Clements:2016de} decomposition.

\subsection{Bottleneck networks}

We now turn to bottleneck quantum communication networks. Indeed, 
using the tools in the previous section we can now fully describe a secret sharing protocol using Gaussian graph states over Gaussian channels which are excellent model for fibre-optic transmission. For the purposes of the discussion in this section it is sufficient consider the case where the sources/detectors are ideal and the only decoherence comes from the lossy channels themselves. This means the total system will be made up of six modes all initialised in the vacuum state. Three modes will be for the tri-partite graph state and three additional modes  $V_A, V_B,V_C$ that will model the corresponding loss channels.

\begin{figure}[htb]
\includegraphics[width=0.45\textwidth]{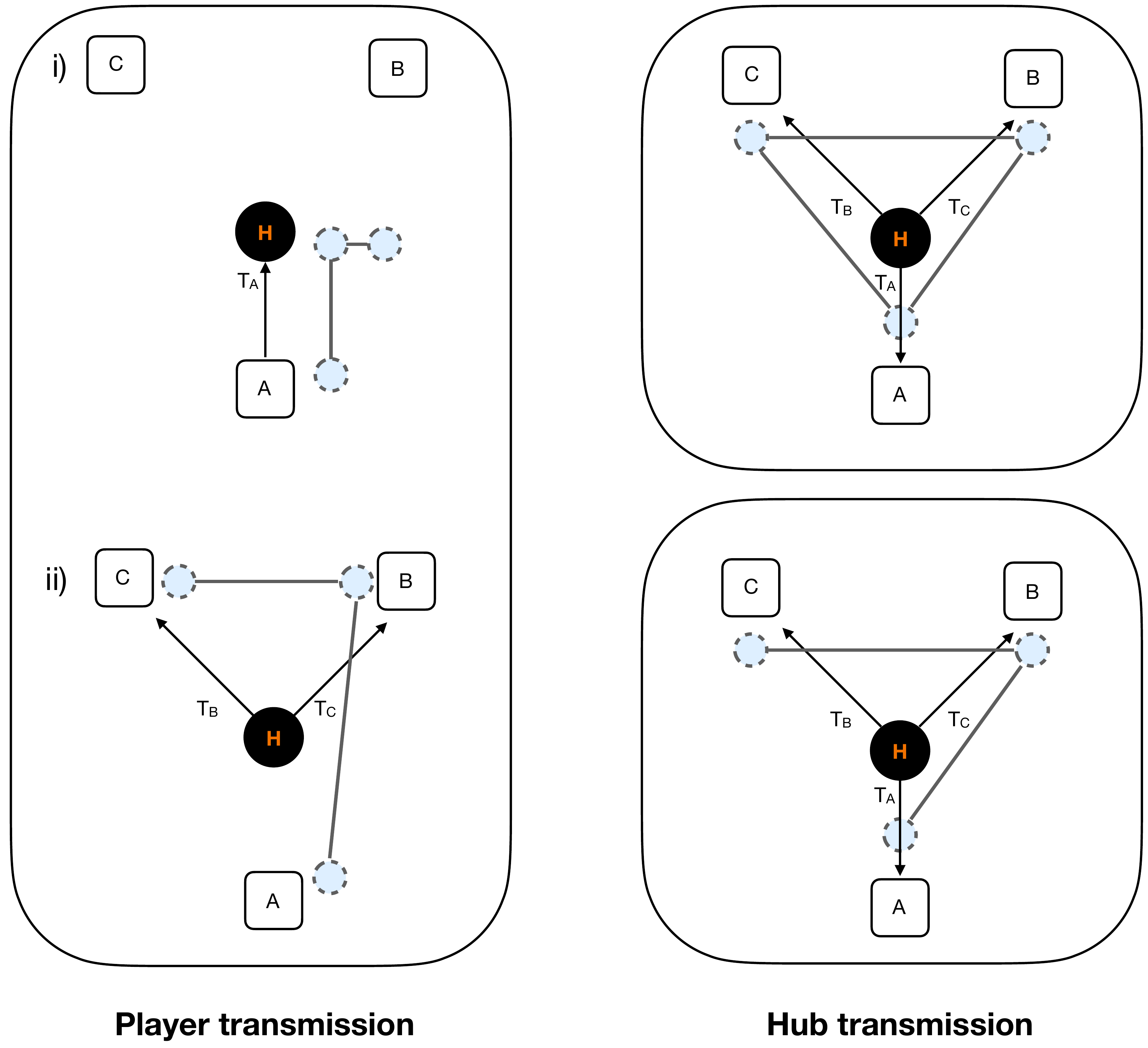}
\caption{Different strategies for distributing graph states through a bottleneck network. The player-in strategy (left) is a two step process: i) Alice makes a two-mode graph state and transmits one mode to the Hub; ii) the Hub entangles this with a third mode, creating a tri-partite graph state, and sends one mode to Bob and the other to Charlie. In the Hub-Out strategy (right) the hub creates the tri-partite graph state directly and transmits one mode to each player in a single step. Unlike the player-in strategy this method can distribute a triangle graph (top) as well as a line (bottom). \label{transstrat}}
\end{figure}

When communicating over a bottleneck network the first noteworthy point is that there are two, inequivalent, network coding strategies that could be employed (Fig.~\ref{transstrat}) to distribute a line graph. The first of these, which we previously denoted the player-in strategy, is where one player first makes a two-mode graph state which is sent to the hub. There it is entangled with a third mode and then all modes are distributed to the corresponding players. The symplectic matrix representing the distribution of the tri-partite line graph over a lossy, bottleneck network is
\begin{eqnarray}
\mathbf{N}_{L} \ee 
\mathbf{BS}_{C,V_C}
(T_C)\cdot\mathbf{BS}_{B,V_B}(T_A)\cdot\mathbf{CZ}_{B,C}(g) 
\nn \\
&\times & \mathbf{BS}_{A,V_A}(T_A)\cdot \mathbf{CZ}_{A,B}(g)\cdot\mathbf{S}_C(-r)\nn \\
&\times & \mathbf{S}_B(-r) \cdot\mathbf{S}_A(-r).
\end{eqnarray}
The second \emph{Hub-Out} strategy involves the creation of the line graph directly at the hub and then distribution. The symplectic matrix for the line graph in a hub out strategy is 
\eqn{\mathbf{N}_{LH} \ee \mathbf{BS}_{C,V_C}(T_C)\cdot\mathbf{BS}_{B,V_B}(T_A)\cdot\mathbf{BS}_{A,V_A}(T_A)\nn \\
&\times &\mathbf{CZ}_{B,C}(g)\cdot \mathbf{CZ}_{A,B}(g)\cdot\mathbf{S}_C(-r) \cdot\mathbf{S}_B(-r) \nn\\
&\times &\mathbf{S}_A(-r)  \label{nlh}.}

Since loss channels and the entangling gates do not commute, these strategies will result in two different states, as can be readily verified by comparing $\mathbf{N}_{L}$ and $\mathbf{N}_{LH}$. Practically speaking, 
there is a significant difference between the two strategies as only in the Hub-Out case can are all three modes in the one location such that we can make use of the simple, offline squeezing preparation method of Fig.~\ref{CV_graph} (iii). For this reason,  we will only consider this strategy for the remainder of the work, but it would be interesting to see what, if any, advantages emerge from the player-in strategy enabled by inline squeezing.

When distributing a line graph, there will be two further possibilities, namely whether the player who is to be the dealer is sent one of the edge nodes of the line or the middle node. Both these possibilities were considered in Fig.~\ref{asympcomp} where we see that the optimal choice is for the dealer to be the middle node. Note that in Eq.~(\ref{nlh}) we have uniquely defined Bob as being the recipient of the central node and hence the optimal dealer. It is also worth noting that with the Hub-Out strategy it is equally possible to prepare a triangle graph as it is a line graph, however our investigations show that this is sub-optimal with respect to the properly chosen line graph.

\subsection{Modelling an experimental implementation \label{expmodel}}
\begin{figure}[htb]
\includegraphics[width=0.45\textwidth]{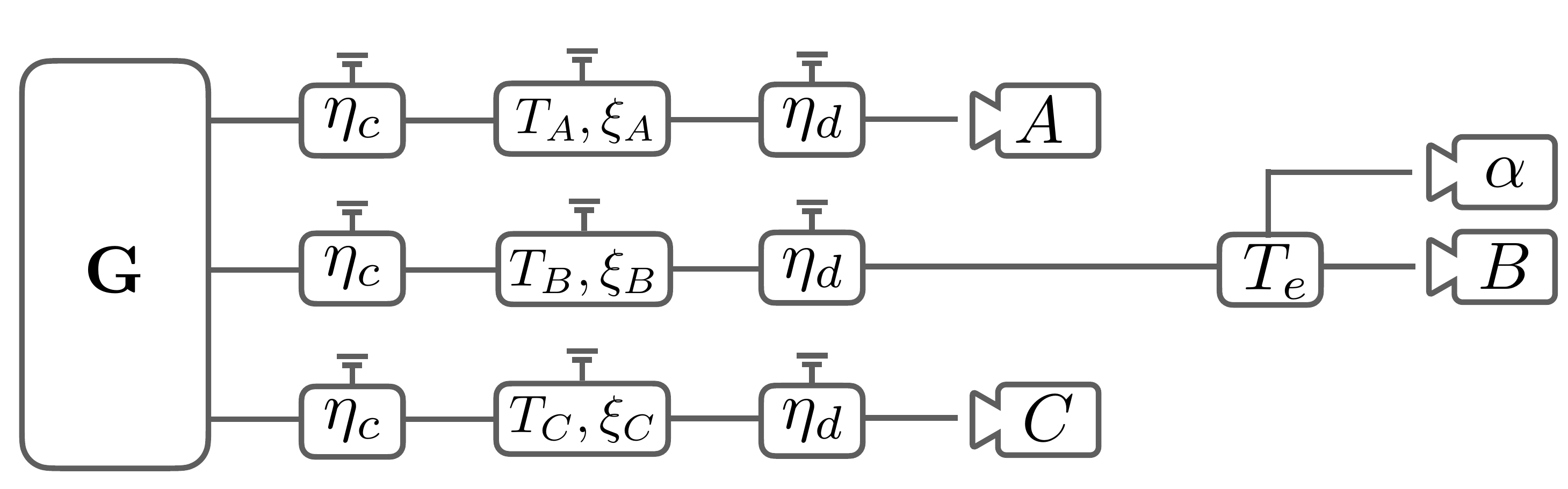}
\caption{Schematic of realistic experimental implementation. Various imperfections are modelled as loss channels with the ground symbol representing lost modes that will be attributed to the dishonest parties. \label{expschem}}
\end{figure}

Whereas our initial, idealised calculations assumed perfect state generation, measurement and transmission through pure lossy channels, in this section we model a more realistic implementation based on past experiments in the literature. A summary of the relevant parameters and their values is given in Table.~\ref{tab:exp} and a schematic of the setup is sketched in Fig.~\ref{expschem}.
\begin{table}[ht]
    \centering
    \begin{tabular}{c|c|c}
      {\bf Symbol}   & {\bf Value} & {\bf Description} \\
      \hline
        $\eta_{\mathrm{es}}$ & 0.99 \cite{Vahlbruch:2016jf} & Escape efficiency\\
        $\eta_\mathrm{f}$ &  0.95 \cite{Gehring:2015ie} & Fibre coupling efficiency \\
         $\eta_d$ &  0.99 \cite{Vahlbruch:2016jf} & Detector efficiency\\
         $r$ & 2.68 (23.3 dB) \cite{Vahlbruch:2016jf} & Inferred squeezing\\
         $T$ & $10^{-0.02 d(\mathrm{km})}$ & Fibre-optic transmission\\
         $\xi$& 0.002 \cite{Jouguet:2013p8197} & Excess noise\\
         \hline 
         
    \end{tabular}
    \caption{Parameters for realistic experimental model.}
    \label{tab:exp}
\end{table}

Turning first to the state generation process, we now consider a finite escape efficiency for the squeezing cavities and a finite coupling efficiency into the optical fibre, which are well modelled by beam-splitters of transmission,  $\eta_{\mathrm{es}}$ and $\eta_{\mathrm{f}}$ respectively mixing the incoming mode with vacuum modes. For simplicity we are taking each squeezer to be identical which means the symmetric loss for the escape efficiency which occurs immediately after squeezing can be commuted through the interferometer in the Bloch Messiah decomposition and combined with the coupling efficiency into a single beam-splitter of transmission $\eta_{\mathrm{c}}=\eta_{\mathrm{es}}\eta_{\mathrm{f}}$. For our calculations we need to infer the initial pure squeezing in Ref.~\cite{Vahlbruch:2016jf}. There a combined total loss of $\eta_{\mathrm{tot}} = 0.975$ is reported along with an \emph{measured} squeezing of 15.3 dB or equivalently a measured squeezed quadrature variance of $V_s = 10^{-15.3/10}$. We can obtain quadrature variance before the loss by inverting $V_s = \eta_{\mathrm{tot}}V_r + 1-\eta_{\mathrm{tot}}$ and finally we find the inferred squeezing parameter of $r = -\log(V_r)/2 = 2.68$ or equivalently 23.3 dB.

Secondly, fibre-optic transmission cannot be completely captured by a pure loss channel. In reality, transmission will induce a small but non-zero excess thermal (and thus Gaussian) noise. This thermal-loss channel can be well modelled as a beam-splitter of transmission $T$ that, instead of mixing the incoming mode with vacuum, combines it with a Gaussian thermal state of variance $1+\xi$. Thirdly, the homodyne detectors will have a finite efficiency which can also be modelled by a lossy beam-splitter of transmission $\eta_d$. Finally, for the purposes of the security proof it is necessary to tap off a small amount of the dealers mode for an energy test. This is done via a another beam-splitter of transmission $T_e$ that reflected a small amount of light to a heterodyne detector (simultaneous measurement of both quadratures) whilst the transmitted mode is homodyned. 

In total this will be a 13 mode system, 4 of which belong to the players ($A$, $B$ and $C$ for the protocol and $B_e$ for Bob's energy test) and the rest which will be attributed to the malicious parties. All are initialised as vacuum states except for the three modes modelling the thermal loss channel which are initialised in a thermal state with variance $1+\xi$. We will label these thermal modes $E_A, E_B, E_C$ as they are assumed to be purified by the eavesdropper. The 6 vacuum modes that are modelling the various losses will be labelled $V_1,\dots ,V_6$. Here we will be taking all detector and coupling efficiencies to be equal. In this notation, the entire realistic model is given by the transform,
\eqn{\mathbf{N}_{\mathrm{exp}} \ee \mathbf{BS}_{C,V_6}(\eta_d)\cdot\mathbf{BS}_{B,V_5}(\eta_d)\cdot\mathbf{BS}_{A,V_4}(\eta_d)\nn \\
&\times&\mathbf{BS}_{B,B_e}(T_e)\cdot\mathbf{BS}_{A,E_C}(T_C)\cdot\mathbf{BS}_{A,E_B}(T_B) \nn \\
&\times &\mathbf{BS}_{A,E_A}(T_A)\cdot\mathbf{BS}_{C,V_3}(\eta_c)\cdot\mathbf{BS}_{B,V_2}(\eta_c)\nn \\
&\times &\mathbf{BS}_{A,V_1}(\eta_c)\cdot\mathbf{CZ}_{A,B}(g)\cdot\mathbf{CZ}_{B,C}(g)\cdot\mathbf{S}_C(-r)\nn\\
&\times &\mathbf{S}_B(-r) \cdot\mathbf{S}_A(-r),}
which is precisely 
reflecting the circuit shown in Fig.~\ref{expschem}. The final covariance matrix is given by
\eqn{\mathbf{\Gamma}_{A,B,C} = \mathbf{N}_{\mathrm{exp}}\cdot\mathbf{N}_{\mathrm{exp}}^\intercal =  \matthree{\mathbf{\Gamma}_A&\mathbf{C}_{A,B}&\mathbf{C}_{A,C}}{\mathbf{C}_{A,B}^\intercal &\mathbf{\Gamma}_B  &\mathbf{C}_{B,C}}{\mathbf{C}_{A,C}^\intercal&\mathbf{C}_{B,C}^\intercal&\mathbf{\Gamma}_C}}
with
\begin{widetext}
\eqn{\mathbf{\Gamma}_A \ee \left(
\begin{array}{cc}
 \eta _d \left(e^{2 r} T_A \eta _c-T_A \left(\xi _A+\eta _c\right)+\xi _A\right)+1 & 0 \\
 0 & T_A \eta _d \left(\eta _c \left(g^2 e^{2 r}+e^{-2 r}-1\right)-\xi _A\right)+\xi _A \eta _d+1 \\
\end{array}
\right) \nn ,\\
\mathbf{\Gamma}_B \ee \left(
\begin{array}{cc}
 \eta _d T_e \left(e^{2 r} T_B \eta _c-T_B \left(\xi _B+\eta _c\right)+\xi _B\right)+1 & 0 \\
 0 & \eta _d T_e \left(T_B \eta _c \left(2 g^2 e^{2 r}+e^{-2 r}-1\right)+\xi _B-\xi _B T_B\right)+1 \\
\end{array}
\right)\nn ,\\
\mathbf{\Gamma}_C \ee \left(
\begin{array}{cc}
 \eta _d \left(e^{2 r} \eta _c T_C-T_C \left(\eta _c+\text{$\xi $C}\right)+\text{$\xi $C}\right)+1 & 0 \\
 0 & T_C \eta _d \left(\eta _c \left(g^2 e^{2 r}+e^{-2 r}-1\right)-\text{$\xi $C}\right)+\text{$\xi $C} \eta _d+1 \\
\end{array}
\right) \nn ,\\
\mathbf{C}_{A,B} \ee \left(
\begin{array}{cc}
 0 & g e^{2 r} \eta _c \eta _d \sqrt{T_e} \sqrt{T_A T_B} \\
 g e^{2 r} \eta _c \eta _d \sqrt{T_e} \sqrt{T_A T_B} & 0 \\
\end{array}
\right)\nn ,\\
\mathbf{C}_{A,C} \ee \left(
\begin{array}{cc}
 0 & 0 \\
 0 & g^2 e^{2 r} \eta _c \eta _d \sqrt{T_A T_C} \\
\end{array}
\right) \nn ,\\
\mathbf{C}_{B,C} \ee \left(
\begin{array}{cc}
 0 & g e^{2 r} \eta _c \eta _d \sqrt{T_e} \sqrt{T_B T_C} \\
 g e^{2 r} \eta _c \eta _d \sqrt{T_e} \sqrt{T_B T_C} & 0 \\
\end{array}
\right).
\label{GExp}}
\end{widetext}

\section{Security analysis for CV secret sharing \label{security}}
We will now present the details of how the secret-fraction in Eq.~(\ref{secfrac}) is derived. 
Essentially, we generalise the proof of Ref.\
\cite{Kogias:2017jz} to the composable, finite-size setting using CVQKD results \cite{Furrer:2012p8365,Furrer:2014eq} which make use of entropic uncertainty relations for the \emph{conditional quantum smooth min- and 
max-entropies} \cite{Furrer:2014ig,Furrer:2011p8368} which are defined as follows. Consider the classical-quantum state of the kind describing the raw measurement registry of the dealer, Alice, and a malicious adversary Eve,
\eqn{\rho_{\mathbf{X}_AE} = \sum_{\mathbf{x}_A} p_{\mathbf{x}_A}\ket{\mathbf{x}_A}\bra{\mathbf{x}_A}\otimes\rho_E^{\mathbf{x}_A} .\label{meas}}
This is the quantum state that is transformed via hashing into the final output described by Eq.\ (\ref{S}). For such a state the conditional quantum  min-entropy is defined as
\eqn{\hmin(\mathbf{X}_A|E)_{\rho_{\mathbf{X}_AE}} = -\log\bk{\sup_{\{E_{\mathbf{x}_A}\}} \sum_{\mathbf{x}_A} p_{\mathbf{x}_A} \mathrm{tr}\bk{E_{\mathbf{x}_A}\rho_E^{\mathbf{x}_A}}}\nn}
where the supremum is taken over all of Eve's possible measurement strategies, i.e., 
her possible POVM's described by operators $\{E_{\mathbf{x}_A}\}$. The corresponding conditional max-entropy is defined as,
\eqn{\hmax(\mathbf{X}_A|E)_{\rho_{\mathbf{X}_AE}} = 2\log\bk{\sup_{\sigma_E}\sum_{\mathbf{x}_A} F(p_{\mathbf{X}_A}\rho_E^{\mathbf{X}_A},\sigma_E)}\nn.}
The smooth versions of these quantities are then given by
\eqn{\hmin^\epsilon(\mathbf{X}_A|E)_{\rho_{\mathbf{X}_AE}} \ee \sup_{\tilde{\rho}_{\mathbf{X}_AE}} \hmin(\mathbf{X}_A|E)_{\tilde{\rho}_{\mathbf{X}_AE}} \nn,\\
\hmax^\epsilon(\mathbf{X}_A|E)_{\rho_{\mathbf{X}_AE}} \ee \inf_{\tilde{\rho}_{\mathbf{X}_AE}} \hmax(\mathbf{X}_A|E)_{\tilde{\rho}_{\mathbf{X}_AE}}, \label{smooth}}
where the supremum and infimum are taken over quantum states that are $\epsilon$-close in the \emph{purified distance}
\begin{equation}
\mathcal{P}(\rho,\sigma)^2 = {1-F^2(\rho,\sigma)}
\end{equation}
where $F(\rho,\sigma)$ denotes the standard fidelity between $\rho$ and $\sigma$. We can now state a crucial result in quantum cryptography, the \emph{leftover hashing lemma} with quantum side-information  \cite{Tomamichel:2011ci,Berta:2011p8367,Furrer:2012p8365,Furrer:2014eq}.

\begin{Lemma}[Leftover hashing lemma]
\label{leftover}
Let $\rho_{\mathbf{X}_AE}$ be a quantum state of the form (\ref{meas}) where $\mathbf{X}_A$ is defined over a a discrete-valued and finite alphabet, E is a finite or infinite dimensional system and $R$ is a register containing the classical information learnt by Eve during information reconciliation. If Alice applies a hashing function, drawn at random from a family of two-universal hash functions that maps $X_A$ to $S_A$ and generates a string of length $\it{l}$, then for any $\epsilon>0$
\eqn{D\bk{\rho_{\mathbf{S}_AE},\tau_{\mathbf{S}_A}\otimes\sigma_E} \leq \sqrt{2^{l - H_{\mathrm{min}}^\epsilon(\mathbf{X}_A|E,R)-2}}+2\epsilon \label{hash}}
where $H_{\mathrm{min}}^\epsilon(\mathbf{X}_A|E,R)$ is the conditional smooth min-entropy of the raw measurement data given Eve's quantum system and the information reconciliation leakage.
\end{Lemma}
Equipped with these tools, we can undertake the security analysis. First, we formalise the arguments behind Eqs.~(\ref{kth1}) and (\ref{lss}).

\begin{Theorem} [Security of $(n,k)$-threshold secret sharing]
For an $(n,k)$-threshold secret sharing protocol as defined in Protocol~\ref{Prot} with trusted and untrusted subsets $T_i\in \mathcal{T}$ and $U_j\in \mathcal{U}$, 
respectively, let $\ell_{\mathrm{EC}}^i$ be the amount of error reconciliation information that would be necessary for the trusted subset $T_i$ and set $\ell_{\mathrm{EC}} = \max_i \ell_{\mathrm{EC}}^i$. A string of length $\ell$ can be extracted that is $(\varepsilon_s + \varepsilon_c)$-secure according to Definition~\ref{SSdef} provided that
\eqn{l  &=&  \min_j \hmin^\epsilon(\bm{X}_A|E,U_j) - \ell_{\mathrm{EC}}^i \nn \\
&-& \log_2\frac{1}{\epsilon_c\epsilon_1^2} +2 .} \label{th1}
\end{Theorem}
\emph{Proof}: The correctness follows straightforwardly from the properties of 2-universal hashing as shown in Refs.\  \cite{Tomamichel:2012p7120,Renner:2005p464}. In step 6 of  Protocol~\ref{Prot}, 
the dealer computes a hash of length $-\log_2\epsilon_c$ chosen uniformly at random from a family of 2-universal hashing functions and transmits the output and the chosen hash-function to all players. When any trusted subset
$T_i$ go to reconstruct the secret they first apply the error correction information, $\ell_{\mathrm{EC}}$, to correct their joint estimate of Alice's string and then use this to evaluate the transmitted hash function. If this is identical to Alice's transmitted hash they proceed otherwise they abort. The necessary correctness is now guaranteed since, by definition, the probability that the two-hashes coincide if there was an error (i.e. if the reconstructed string differs from Alice's string) is at most $2^{\lceil \log_2\epsilon_c\rceil} \leq \epsilon_c$. In order to actually compute the achievable performance we will ultimately need to quantify how large $\ell_{\mathrm{EC}}^i$ must be in order for a given subset to successfully correct their string and pass the correctness check with high probability and we will explain this in the next section. However, for the purposes of the security proof, however $\ell_{\mathrm{EC}}^i$ is chosen, passing the hashing check ensured the $\epsilon_c$-correctness of the conditional output. Moreover, taking the worst over all $T_i$ ensures the correctness holds for all trusted subsets, thereby satisfying the correctness condition in Definition~\ref{SSdef}. In the worst-case where Eve learns one bit of the key for every bit announced during error reconciliation we have that 
\eqn{\min_i \hmin^\epsilon(\mathbf{X}_A|E,U_j,R_i) &\geq& \hmin^\epsilon(\mathbf{X}_A|E,U_j) \nn \\
&-& \max_i\ell^i_{\mathrm{EC}} - \log_2\frac{1}{\epsilon_c}.}
The secrecy is a straightforward consequence of the leftover hashing lemma. Considering Eq.~(\ref{hash}) and redefining the eavesdropper system to include the $j^{\mathrm{th}}$ untrusted subset, $U_j$, we can see that by choosing 
\eqn{l \ee \hmin^\epsilon(\mathbf{X}_A|E,U_j,R) + 2 -2 \log \frac{p_{\mathrm{pass}}}{\epsilon_1}} for some $\epsilon_1>0$ then the right hand side becomes $\epsilon_1/p_{\mathrm{pass}}+2\epsilon$. Choosing $\epsilon = (\epsilon_s - \epsilon_1)/(2p_{\mathrm{pass}})$ and substituting in Eq.~(\ref{hash}) gives
\eqn{D\bk{\rho_{\mathbf{S}_AEU_j},\tau_{\mathbf{S}_A}\otimes\sigma_{EU_j}} \leq \frac{\epsilon_s}{p_{\mathrm{pass}}}.} 
 Putting this together and using the fact that $\log_2 p_{\mathrm{pass}}<0$ means that for the unauthorised subset $U_j$ a hashing to a key of length
\eqn{l  &=&  \hmin^\epsilon(\bm{X}_A|E,U_j) - \max_i \ell_{\mathrm{EC}}^i \nn \\
&-& \log_2\frac{1}{\epsilon_c\epsilon_1^2} +2,}
will ensure that
\eqn{p_{\mathrm{pass}}D\bk{\rho_{\mathbf{S}_AEU_j},\tau_{\mathbf{S}_A}\otimes\sigma_{EU_j}} \leq \epsilon_s.}
If the length is chosen by taking the minimum of $\hmin^\epsilon(\bm{X}_A|E,U_j)$ over all untrusted subsets then the secrecy condition in Definition~\ref{SSdef} is immediately satisfied which completes the proof. $\qed$

All that remains is to find a way to bound the min-entropy for each $U_j$, whilst avoiding any participant attacks. The key insight is that, for any fixed $U_j$, the cryptographic situation is identical to a QKD protocol where the roles of Eve and Bob are played by $U_j$ the corresponding complementary set $C_j$. Security for a realistic CV protocol can then be established via the results of Furrer \cite{Furrer:2014eq}. Protection against participant attacks is now guaranteed since the parameter estimation steps exclude the untrusted subset so there is no opportunity for them to cheat by manipulating the observed statistics as suggested in Ref.\ \cite{Karlsson:1999ez}. For completeness we state the necessary theorem in full and sketch the proof, highlighting the point at which the standard participant attacks would occur in a less careful analysis. For a full proof, 
we refer the reader to Ref.\  \cite{Furrer:2014eq}.

\begin{Theorem}[Adapted from
Ref.\ \cite{Furrer:2014eq}]
For an ($n,k,m,t,p$) secret sharing protocol as defined in Protocol~\ref{Prot}, carried out with coarse-grained quadrature measurements of resolution $\delta_{\mathbb{X},\mathbb{P}}$ and maximum value $M$, the conditional smooth min-entropy  $\hmin^\epsilon(\bm{X}_A|E,U_j)$ of the subset $U_j$ in collaboration with Eve, conditioned on passing a correlation threshold $d_0^j$ with the complementary set $C_j$ and an additional $(T_e,\alpha)$ is lower bounded by
\eqn{m\left[\log _{2} \bk{\frac{1}{c(\delta_{\mathbb{X}},\delta_{\mathbb{P}})}}-\max_j \log _{2}\bk{ \gamma\left(d_{0}^j+\mu\right)}\right] \label{secfracapp}}
where 
\begin{equation}
\gamma(d)=\left(d+\sqrt{1+d^{2}}\right)\left(\frac{d}{\sqrt{1+d^{2}}-1}\right)^{d} \label{gamma}
\end{equation}
and
\eqn{\mu=\sqrt{2 \log _{2} \epsilon_\mu^{-1}} \frac{N \sigma_{*}}{t \sqrt{m}}+\frac{4(M / \delta) \log _{2} \epsilon_\mu^{-1}}{3} \frac{N}{m t} \label{mu}}
with 
\eqn{\begin{aligned}
\sigma_{*}^{2}=& \frac{t_j}{N}\left(V_{d}^{\mathrm{PE}}-\frac{t_j}{N}d_0^{2}\right)+\frac{m}{N}\left(V_{A}^{\mathrm{PE}}+V_{B}^{\mathrm{PE}}+2 \frac{v}{\delta_{\mathbb{P}}^{2}}\right) 
\end{aligned}
\label{sig}}
for the smallest $v$ for which
\eqn{\begin{aligned}
\epsilon_\mu=&\left(\epsilon_{s}-\epsilon_{1}-2 \sqrt{2 m \Gamma(M, T_e, \alpha)}\right)^{2} \\
&-2 \exp \left(-2(v / M^2)^{2} \frac{m t_j^{2}}{(N)(t_j+1)}\right)\label{epmu}
\end{aligned}}
is positive and 
\begin{equation}
\epsilon_{s}-\epsilon_{1}-2 \sqrt{2 m \Gamma(M, T_e, \alpha)}>0 
\end{equation}
also holds (if either of these positivity conditions cannot be satisfied, the secret fraction is actually 
zero). Here $N=m+t_j$, $V_{P_{A}}^{\mathrm{PE}}$ and $V_{P_{B}}^{\mathrm{PE}}$ are the observed variances of the $\mathbb{P}$ measurements used for parameter estimation, $V_{d}^{\mathrm{PE}}$ is the variance of their absolute difference $|P_A-P_B|$ and
\eqn{\Gamma(M, T_e, \alpha)&:=&\frac{\sqrt{1+\lambda}+\sqrt{1+\lambda^{-1}}}{2}\nn \\
&&\times\exp \left(-\frac{(\zeta M-\alpha)^{2}}{T(1+\lambda) / 2}\right)\label{Gamma}}
with $\zeta = \sqrt{(1-T)/(2T)}$ and $\lambda = ((2T-1)/T)^2$.
\end{Theorem}
\emph{Proof sketch:} For a single fixed $U_j$ and corresponding $C_j$, Protocol 1 becomes identical to CVQKD protocol of Ref.\  \cite{Furrer:2014eq} if we identify the systems $C_j:=B$ as single entity, Bob, and $U_jE:=E$ as a single adversary Eve. We can apply the CVQKD proof which we now sketch. The basic idea is to use an 
\emph{entropic uncertainty relation (EUR)} for the smooth min- and max-entropies. For realistic measurements with a finite range, the corresponding EUR becomes trivial. However, for a sufficiently tight upper bound on the energy of the incoming state, the resultant smooth min- and max entropies can be rigorously related to those of an ideal, infinite-range, measurement for which there is a useful EUR. Such an upper bound is precisely what is achieved by the additional energy test carried out via heterodyne detection. Statistical deviation bounds can then be applied to turn the observed correlations in the certification basis into a guarantee for the smooth min-entropy of the key generation measurements.

\emph{Step 1. Entropic uncertainty relation.} A realistic quadrature measurement with a resolution $\delta$ and finite detection window $[-M,M]$ can be represented as a series of projections in the corresponding basis onto intervals 
\eqn{
I_{1} &=&(-\infty,-M+\delta]\nn ,\\
I_{k} &=&(-M+(k-1) \delta,-M+k \delta],\nn \\
I_{2 M / \delta} &=&(M-\delta, \infty),\label{I} }
with $k=2, \ldots, 2 M / \delta-1$ and where the finite range is captured by the semi-infinite end bins. The measurement $\mathbb{X}_A = \{E^{\mathbb{X}_A}_i\}$ is given by measurement operators
\eqn{E^{\mathbb{X}_A}_i = \int_{I_i} q_A \ket{q_A}\bra{q_A} dq_A \label{quad}, \hs q_A\in \{x_A, p_A\},}
where the intervals are defined according to Eq.~(\ref{I}) with resolutions $\delta_{\mathbb{X}}$. The measurements $\mathbb{P}_B = \{E^{\mathbb{P}_B}_i\}$ etc are defined analogously. We allow $\delta_{\mathbb{P}}$ and $\delta_{\mathbb{X}}$ to differ but for simplicity will assume that each quadrature resolution is symmetric for all parties and we will take the range $M$ to be the same for all measurements. 

By contrast, an infinite-range measurements ($\tilde{\mathbb{X}},\tilde{\mathbb{X}}$) with the same resolution would be described by the projections onto the intervals
\eqn{\tilde{I}_{k}=(M+(k-1) \delta,-M+k \delta], \quad k \in \mathbb{Z}.}
These infinite range measurements can be shown to give rise to the following EUR for a joint state vector $\ket{\Psi_{ABE}}$,
\begin{equation}
\hmin^\epsilon\left(\bm{\tilde{X}}_{A}^{\mathrm{key}} | E\right) \geqslant mq(\mathbb{\tilde{X}}, \mathbb{\tilde{P}} ) -\hmax^\epsilon\left(\bm{\tilde{P}}_{A}^{\mathrm{key}} |B\right),\label{eurapp}
\end{equation}
where we have made the number of measurements $m$ -- and the fact that we are specifically interested in the $\mathbb{X}$ measurements that were used to generate a secret key -- explicit. For this setup, 
it has been shown that \cite{Furrer:2014ig,Berta:2011p8367}
\eqn{q(\mathbb{\tilde{X}}, \mathbb{\tilde{P}}) = -\log_2\bk{ \frac{\delta_{\mathbb{X}}\delta_{\mathbb{P}}}{2 \pi} \cdot S_{0}^{(1)}\left(1, \frac{\delta_{\mathbb{X}}\delta_{\mathbb{P}}}{4}\right)^{2} }}
with $S_n^{(1)}(\cdot,u)$ the radial prolate spheroidal wave function of the first kind and is related to the complementarity of the . Since $q(\mathbb{\tilde{X}}_A, \mathbb{\tilde{P}}_A) $ is positive, for sufficiently good $\mathbb{\tilde{P}}$ correlations between Alice and Bob this will give a useful bound on the conditional min-entropy. Unfortunately, for finite measurements the semi-infinite end bins have significant overlap and we find that $q(\mathbb{X}_A, \mathbb{P}_A) \approx 0$ and the EUR becomes trivial.
Intuitively, we would expect that for states that have a support lying almost entirely inside the range $[-M,M]$ in both quadratures that the difference between using a finite or infinite range detector should be operationally negligible for all quantities, including the smoothed entropies. This intuition can be made rigorous if we have a bound on the purification distance, which appears in the definitions in Eq.~(\ref{smooth}), between the post-measurement states, that would arise from finite or infinite range measurements. Concretely, given a promise that
\eqn{\mathcal{P}\bk{\rho_{\mathbf{X}_ABE},\rho_{\mathbf{\tilde{X}}_ABE}} &\leq& \tilde{\epsilon}\nn ,\\ 
\mathcal{P}\bk{\rho_{\mathbf{P}_ABE},\rho_{\mathbf{\tilde{P}}_ABE}} &\leq& \tilde{\epsilon},}
it can be shown that
\eqn{\hmin^\epsilon\left(\bm{X}_{A} | E\right) &\geq& \hmin^{\epsilon- \tilde{\epsilon}}\left(\bm{\tilde{X}}_{A} | E\right) \nn ,\\
\hmax^\epsilon\left(\bm{\tilde{P}}_{A} | E\right) &\leq& \hmax^{\epsilon- \tilde{\epsilon}}\left(\bm{P}_{A} | E\right). \label{pure_bound}}
In combination with Eq.~(\ref{eurapp}) this yields
\begin{equation}
\hmin^\epsilon\left(\bm{X}_{A}^{\mathrm{key}} | E\right) \geqslant mq(\mathbb{\tilde{X}}, \mathbb{\tilde{P}} ) -\hmax^{\epsilon-2\tilde{\epsilon}}\left(\bm{P}_{A}^{\mathrm{key}} |B\right). \label{hminbound1}
\end{equation}
This is now the kind of relationship we
have intended to find, where the entropies of the realistic measurements ($\mathbb{X}, \mathbb{P} $) are related to one another and the positive, and hence useful, entropic constant of the infinite-range measurements ($\mathbb{\tilde{X}}, \mathbb{\tilde{P}} $). Next we require a way to bound $\tilde{\epsilon}$.

\emph{Step 2. Energy test.} Using a beam-splitter with known transmission $T_e$ to mix the incoming state with a trusted vacuum mode it is possible to tap off a small amount of the incoming light for analysis. Our goal is to bound the purified distance between states measured with either finite, or infinite range measurements as per Eq.~(\ref{pure_bound}). Unsurprisingly, it turns out this quantity can be bounded as long as we have a restriction on the probability that a detection outside the range $[-M,M]$ would ever occur during the protocol. Concretely, it can be shown that for a protocol with a total of $N$ rounds that
\eqn{\mathcal{P}\left(\rho_{\mathbf{X}_{A} BE}, \rho_{\mathbf{\tilde{X}}_{A} BE}\right)^2 \leqslant {1-\operatorname{Pr}\left[\wedge_{i}\left\{\left|q_{i}\right| \leqslant M\right\} \mid \rho_{A^{m} B^{m} E}\right]^{2}}\nn}
where $\left\{\left|q_{i}\right| \leqslant M\right\}$ denotes the event that the absolute value of a continuous $\hat{x}$-quadrature measurement of Alice’s $i^{\mathrm{th}}$ mode is smaller than $M$ and a corresponding result for $\mathcal{P}\left(\rho_{\mathbf{P}_{A} BE}, \rho_{\mathbf{\tilde{P}}_{A} BE}\right)$.
This probability can be estimated from the tapped off beam. Since we would like to bound the probability for both quadratures it is necessary to perform a heterodyne detection - mixing the tapped of light with a further vacuum mode on a balanced beam-splitter and then measuring the $x$ quadrature on one output and $\hat{p}$ on the other. We will say that the energy test is passed if, for all rounds, neither quadrature value exceeded some threshold $\alpha$. Defining $p_{\mathrm{pass}}$ as the probability that the test it can then be shown that, conditioned on passing that both $\mathcal{P}\left(\rho_{\mathbf{X}_{A} E}, \rho_{\mathbf{\tilde{X}}_{A} E}\right)$ and $\mathcal{P}\left(\rho_{\mathbf{P}_{B} E}, \rho_{\mathbf{\tilde{X}}_{B} E}\right)$
\eqn{ \tilde{\epsilon}^2 = {\frac{2 m \Gamma(M, T_e, \alpha)}{p_{\text {pass }}},} \label{eptilde}}
where the function $\Gamma:\mathbbm{R}^+\times\mathbbm{R}^+
\times \mathbbm{R}^+\rightarrow \mathbbm{R}^+$
 is given by Eq.~(\ref{Gamma})

\emph{Step 3. Statistical bounds on the max-entropy.}
As mentioned before and EUR describes a \emph{counterfactual} situation. For example, in our case, 
Eq.~(\ref{hminbound1}) lower bounds $\hmin^\epsilon(\bm{X}_{A}^{\mathrm{key}} | E)$, Eve's smooth min-entropy regarding Alice's $m$ key generation measurements, and $\hmax^{\epsilon-2\tilde{\epsilon}}(\bm{P}_{A}^{\mathrm{key}} |B)$, the max-entropy of Bob regarding Alice's measurements \emph{if she had instead chosen to measure with $\mathbb{P}$ for those rounds}. Strictly speaking, we have no direct access to this latter quantity  since, by definition, Alice measures with $\mathbb{X}$ rather than $\mathbb{P}$ for those rounds. What we have instead are the strings $\bm{P}_A^{\mathrm{PE}}$ and $\bm{P}_B^{\mathrm{PE}}$ arising from the $t$ parameter estimation rounds where Alice and Bob both measured in $\mathbb{P}$. However, provided these rounds were chosen randomly, then these observed correlations can be used to give a rigorous, probabilistic bound on what the correlations between $\bm{P}_A^{\mathrm{key}}$ and $\bm{P}_B^{\mathrm{key}}$ would have been. These can in turn bound Bob's max-entropy in Eq.~(\ref{hminbound1}).

This is the precise point at which the potential for participant attacks formally enters in the security analysis. The situation we find ourselves in is the so-called \emph{sampling without replacement} scenario and there is an entire machinery of large deviation bounds that use the observed statistics of a randomly chosen sample to probabilistically bound the behaviour of the remaining population. However, all of these results are only valid in the case that the $t_j$ parameter estimation rounds were truly chosen at random and constitute a fair sample. In a QKD protocol, or equivalently for a known, fixed untrusted subset of a secret sharing protocol then this condition is automatically satisfied since the probability of a round being used for parameter estimation is determined solely by Alice (equivalently the dealer) and Bob (equivalently the complementary subset) who are trusted by definition. As explained earlier, Theorem~\ref{th1} shows that by minimising the key rate over all such tests, security of the overall threshold scheme is guaranteed. This is in contrast to the original HBB scheme \cite{Hillery:1999tb} where every check round is dependent on the basis choice of all players, thereby always including parties who by definition cannot be trusted. Deviation bounds are therefore not valid and security cannot be established. Indeed, it is precisely this problem of malicious participants biasing the parameter estimation statistics that has been exploited in the original works demonstrating attacks that can compromise or completely break the security of the HBB scheme \cite{Karlsson:1999ez,Qin:2007fv}.

For the purposes of this theorem, the selection of rounds is determined solely by parties that can be assumed honest and so fair sampling is assured and we can proceed. The first tool is the result that for any $\varepsilon$, provided \eqn{\mathrm{Pr}[d(\bm{P}_A^{\mathrm{key}},\bm{P}_B^{\mathrm{key}}) >d_{\mathrm{key}}]\leq \varepsilon^2 \label{pr_dk}} 
it can be shown that \cite{Furrer:2012p8365},
\eqn{\hmax^{\varepsilon}(\bm{P}_{A}^{\mathrm{key}} |B) < m \log_2(\gamma(d_{\mathrm{key}})),\label{hmaxbnd}}
with $\gamma:\mathbbm{R}\rightarrow \mathbbm{R}$ being given by Eq.~(\ref{gamma}). Obtaining a bound such as Eq.~(\ref{pr_dk}) can be achieved using large deviation bounds. Recall that the parameter estimation test is only passed if the strings generated in the $t_j$ parameter estimation rounds satisfy
\eqn{d(\bm{P}_A^{\mathrm{PE}},\bm{P}_B^{\mathrm{PE}}) \leq d_0.}
Denoting ($\bm{P}_A^{\mathrm{tot}},\bm{P}_B^{\mathrm{tot}}$) as the total results from all $N=m+t_j$ rounds it must be the case that 
\eqn{Nd_{\mathrm{tot}} = md_{\mathrm{key}} + t_jd_0 \label{dtot}} where $d_{\mathrm{tot}} = d\bk{\bm{P}_A^{\mathrm{tot}},\bm{P}_B^{\mathrm{tot}}}$ is the total distance over all rounds. 

Hoeffding's application of Bernstein's inequality to the replacement without sampling case can then be used with  Eq.~(\ref{dtot}) to show that
\eqn{\operatorname{Pr}\left[d_{\mathrm{key}} \geqslant d_{0}+\mu|\sigma \right] \leqslant \epsilon_\mu}
with
\eqn{\epsilon_\mu = \exp \left(-\frac{n \mu^{2}(t_j/N)^2}{2 \sigma^{2}+2 \mu(t_j/N)(M/\delta_{\mathbb{P}}) / 3}\right) \label{epmudef}}
where we have conditioned on knowing the average variance of the total population, $\sigma^2 = (1/N)\sum_
{i=1}^N |\bm{P}_A^{\mathrm{tot}}-\bm{P}_B^{\mathrm{tot}}|^2 - d_{\mathrm{tot}}^2$. However, $\sigma$ itself also needs to be bounded from the observed statistics. We show this calculation in some more detail as we both slightly improve on, and correct some small errors in Ref.~\cite{Furrer:2014eq}. We have that
\eqn{\sigma^2 \ee \frac{1}{N}\sum_
{i=1}^N |\bm{P}_A^{\mathrm{tot}}-\bm{P}_B^{\mathrm{tot}}|^2 - d_{\mathrm{tot}}^2 \nn \\
\ee \frac{1}{N}\bk{\sum_{i=1}^{t_j} |\bm{P}_A^{\mathrm{PE}}-\bm{P}_B^{\mathrm{PE}}|^2 + \sum_{i=1}^m |\bm{P}_A^{\mathrm{key}}-\bm{P}_B^{\mathrm{key}}|^2} \nn \\
&-& \bk{\frac{t_j}{N}d_0}^2 - \bk{\frac{m}{N}d_{\mathrm{key}}}^2 - \frac{2mt}{N^2}d_0d_{\mathrm{key}}}
where we have split up the sum and have used Eq.~(\ref{dtot}). We can drop the last two negative terms involving $d_{\mathrm{key}}$ and recalling that 
\eqn{V_d^{\mathrm{PE}} \ee \frac{1}{t_j}\sum_{i=1}^{t_j}|\bm{P}_A^{\mathrm{PE}}-\bm{P}_B^{\mathrm{PE}}|^2 \nn, \\
V_A^{\mathrm{PE}} \ee \frac{1}{t_j}\sum_{i=1}^{t_j}|\bm{P}_A^{\mathrm{PE}}|^2, V_B^{\mathrm{PE}} = \frac{1}{t_j}\sum_{i=1}^{t_j}|\bm{P}_B^{\mathrm{PE}}|^2 \nn ,\\
V_A^{\mathrm{key}} \ee \frac{1}{m}\sum_{i=1}^m|\bm{P}_A^{\mathrm{key}}|^2,\hs  V_B^{\mathrm{key}} = \frac{1}{m}\sum_{i=1}^m|\bm{P}_B^{\mathrm{key}}|^2 \nn,} 
we have that
\eqn{\sigma^2 &\leq& \frac{t_j}{N} V_d^{\mathrm{PE}} + \frac{1}{N}\sum_
{i=1}^m |\bm{P}_A^{\mathrm{key}}|^2 + \frac{1}{N}\sum_
{i=1}^m|\bm{P}_B^{\mathrm{key}}|^2\nn \\
&-& \bk{\frac{t_j}{N}d_0}^2  \nn\\
\ee \frac{t_j}{N}\bk{V_d^{\mathrm{PE}} -\frac{t_j}{N}d_0^2}+ \frac{m}{N} \bk{V_A^{\mathrm{key}}+V_B^{\mathrm{key}} } }
where in the first inequality we also have used that $|x-y|^2\leq |x|^2 + |y|^2$. Although $V_A^{\mathrm{key}}$ and $V_B^{\mathrm{key}}$ are not directly observed they can be constrained via Serfling's bound \cite{Serfling:1974dx} which can be used to show that
\eqn{\operatorname{Pr}\left[V_A^{\mathrm{key}} \geqslant V_A^{\mathrm{PE}}+\frac{v}{\delta_{\mathbb{P}}^2}\right] \leqslant \epsilon_v}
where
\eqn{\epsilon_v = \exp \left(\frac{-2 v^{2} m t_j^{2}}{M^{4}(m+t)(t_j+1)}\right)}
with the analogous result holding for $V_A^{\mathrm{key}}$. 
This means that $\sigma$ is upper bounded by $\sigma^*$ as given in Eq.~(\ref{sig}) except with probability $2\epsilon_v$. 
In total, we have that 
\eqn{\operatorname{Pr}\left[d_{\mathrm{key}} \geqslant d_{0}+\mu\right] \leqslant \epsilon_\mu + 2\epsilon_v.}
Applying Eq.~(\ref{hmaxbnd}) to Eq.~(\ref{hminbound1}),
we see that to bound the max-entropy and hence the final key length we require,
\eqn{(\epsilon_\mu + 2\epsilon_v)^2 = \frac{\epsilon_s - \epsilon_1}{2p_{\mathrm{pass}}} -2 \tilde{\epsilon}}
where $\tilde{.}$ is given by Eq.~(\ref{eptilde}). This is precisely equivalent to Eq.~(\ref{epmu}), though we must always ensure that the RHS is positive and $\epsilon_\mu$ are positive. Solving Eq.~(\ref{epmudef}) for the necessary $\mu$ gives Eq.~(\ref{mu}) which completes the proof. \qed 

\section{Evaluating the secret fraction \label{evaluate}}
In this section, 
we explain how to evaluate equations (\ref{kasymp}) and (\ref{secfrac}), as well as the relevant QKD comparisons shown in Fig.~\ref{asympcomp}, Fig.~\ref{advantangeregion} and Fig.~\ref{finitecomp}. We also present some additional key rate results and also the values of the parameters that have been optimised over to produce the figures.

\subsection{Idealised, asymptotic results \label{asympapp}}
In the main text briefly explained how the standard asymptotic key rate arise from the composably finite-sized secret fraction in the limit of infinitely long key exchange and perfect equipment. To recapitulate, starting from Eq.~(\ref{lss}) we can write the secret fraction for a fixed $T_i$ and $U_j$ as,
\eqn{\frac{l}{L} \ee \frac{1}{L}\bk{\hmin^\epsilon(\bm{X}^m_A|E,U_j) - \ell_{\mathrm{EC}}^i - \log_2\frac{1}{\epsilon_c\epsilon_1^2} +2 }\nn.}
In the asymptotic limit, is has been shown that collective eavesdropping attacks are optimal \cite{Renner:2009p1}, hence we can assume that the state is of the form $\rho_{ABE}^{\otimes m}$. Then, the asymptotic equi-partition theorem for infinite dimensions states that,
\eqn{\lim_{m\rightarrow\infty} \frac{1}{m}\hmin^\epsilon(\bm{X}^m_A|E,U_j) = S(X_A|E,U_j)}
where the is the conditional von Neumann entropy of Alice's measurement $X_A$ given $EU_j$ defined in Eq.~(\ref{cvn}). Shannon's noisy coding theorem says that asymptotically, since we are free to assume a i.i.d. structure in the worst case, with ideal error reconciliation we have that
\eqn{\ell_{\mathrm{EC}}^i = m H(X_A|X_{T_i}).} 
A critical point is the value of $m,t$ and $t_j$. If the probability for measuring in the key quadrature is $p$ then, asymptotically, for an $(n,n)$-threshold protocol we would have that 
\eqn{m=p^{n+1}L, \hs t_j = (1-p)^2L \label{mt}}
This is because for a valid key generation round we need all three parties to measure in the key basis, and to be a useful parameter estimation round both the dealer and all players in one of the complementary subsets, $C_j$, must have chosen the check basis. For an $(n,n)$ scheme each $U_j$ has $n-1$ players so each $C_j$ consists of just one player. This means that the total number of valid parameter estimation rounds in the probability that the dealer and at least one player both measure in the check basis yielding
\eqn{t = (1-p)(1-p^{n-1})L. }
Thus a total of $N=m+t$ rounds out of the total $L$ rounds are used for either key generation or parameter estimation with the remaining rounds discarded. However, in the asymptotic limit the protocol becomes arbitrarily efficient. In the limit of infinite data then we can still acquire perfect parameter estimation statistics by sacrificing an arbitrarily small proportion of data since for any $p$ in the limit $L\rightarrow \infty$ both $m$ and $t$ also tend to infinity. Thus we can take the limit $p\rightarrow 1$ which in turn means $m\rightarrow L$. Taken together, this gives
\begin{eqnarray}
\lim_{L\rightarrow\infty}\frac{l}{L} \ee S(X_A|X_E,U_j) - H(X_A|X_{T_i}) \nn\\
\ee I(X_A:X_{T_i}) - \chi(X_A:E,U_j)
\end{eqnarray}
where in the second line we have used the \emph{mutual information} 
 \eqn{I(X_A:Y) := H(X) - H(X|Y)\label{iabapp}} and the \emph{Holevo quantity} 
 \begin{equation}
 \chi(X_A:E) := S(E) - \sum_{x_A} p(x_A)S(E|x_A), \label{holevoapp}
 \end{equation} 
 to rewrite in the form more commonly found in the CVQKD literature. We can see this analysis is tight since this expression coincides with the asymptotically optimal Devetak-Winter rate \cite{Devetak:2005p5086}. The secret sharing rate can then be calculated by taking the worst case for $T_i$ and $U_j$, which in fact recovers Eq.~(\ref{kasymp}).

 To evaluate Eq.~(\ref{holevoapp}), 
 it suffices to recall that, by definition, the joint state vector $\ket{\Psi_{AC_jU_jE}}$ represents a pure quantum state, which means we know that $S(E,U_j) = S(A,C_j)$ and $S(E,U_j|x_A) = S(C_j|x_A)$. Computing these expressions can be dramatically simplified by first noting that, for ideal measurement devices making perfect quadrature measurements the entire protocol, including the conditional states would be perfectly Gaussian in the absence of an eavesdropper. Secondly, it has been shown that, asymptotically, it is optimal for an eavesdropper that the final state also be Gaussian \cite{Navascues:2006p805,GarciaPatron:2006p381}. In combination, 
 this with the fact that, asymptotically, it has been shown that the Gaussian attacks are optimal and we can safely assume that the final state is entirely Gaussian. In this case the von Neumann entropy is solely a function of the relevant covariance matrix \cite{Weedbrook:2012p5160}. For an $N$-mode state on can then write
 \eqn{S(\rho) = S(\mathbf{\Gamma})= \sum_{i=1}^N g(\lambda_i)}
 where $g:[0,\infty)\rightarrow [0,\infty)$ is defined as
 \eqn{g(x) = \bk{\frac{x+1}{2}}\log_2\bk{\frac{x+1}{2}} - \bk{\frac{x-1}{2}}\log_2\bk{\frac{x-1}{2}}\nn}
 and where the $\{\lambda_k\}$ again are the \emph{symplectic eigenvalues} of the corresponding covariance matrix $\mathbf{\Gamma}$, which are 
 defined by the (singly counted) eigenvalues of the matrix $|i\mathbf{\Omega\Gamma}|$. 

Note that this is different from the strategy adopted by Ref.~\cite{Kogias:2017jz}, where they instead bounded the malicious parties information (or equivalently their conditional entropy) via and EUR for the asymptotic von Neumann entropies and ideal quadrature measurements \cite{Furrer:2014ig}
 \eqn{S(X_A|E,U_j) + S(P_A|C_j) \geq \frac{1}{2\pi}.}
This approach could be thought of as first applying the finite-size min- and max-entropy EUR of Eq.~(\ref{eur}) which holds without any assumptions and then taking the asymptotic limit and invoking the asymptotic optimality of collective attacks. However, as explained in the main text, this is unnecessary and will result in pessimistic estimates of the key rate due the looseness of the EUR in the relevant case \cite{Furrer:2014eq,Walk}. Therefore, in the asymptotic setting the key rates for secret sharing in this work are higher than those found in Ref.~\cite{Kogias:2017jz}.

Turning to our concrete situation of a $(2,2)$-scheme using Hub-Out transmission over a lossy bottleneck network, we can now compute everything given the output CM. For an ideal system, where the only imperfections are the losses from the channel transmission the final CM is given by 
\eqn{\mathbf{\Gamma}_{A,B,C} = \mathbf{N}_{LH}\cdot \mathbf{N}_{LH}^\intercal \label{GABC}}
with $\mathbf{N}_{LH}$ given by Eq.~(\ref{nlh}). Equivalently, 
one could take Eq.~(\ref{GExp}) in the limit of perfect implementation ($\eta_d=\eta_c = T_e=1$, $\xi=0$). For example, assuming an honest Bob and Alice encoding her key in the $\hat{x}$ basis, the Holevo information of a dishonest Charlie collaborating with Eve is
\eqn{\chi(X_A:EC) = S(AB) - S(B|x_A).}
The CM $\mathbf{\Gamma}_{A,B}$ is simply the appropriate sub-matrix of Eq.~(\ref{GABC}) and the conditional CM $\mathbf{\Gamma}_{B|x_A}$ can be obtained from $\mathbf{\Gamma}_{A,B}$ via the Schur complements in Eq.~(\ref{schur}). Notice that, in comparison to Eq.~(\ref{holevoapp}), there is now no sum over $x_A$. This is because the conditional CM is independent of the actual value of $x_A$. 

For Gaussian distributions the mutual information between Alice's $\hat{x}$ measurement and Bob's $\hat{p}$ is given by
\eqn{I(X_A:P_B) = \frac{1}{2}\log_2\bk{\frac{V_{X_A}}{V_{X_A|P_B}}}}
where $V_{X_A}$ and $V_{X_A|P_B}$ are given by the first entry of $\mathbf{\Gamma}_{A}$ and $\mathbf{\Gamma}_{A|p_B}$ respectively. To get the secret sharing rate we have to carry out the same computation for an untrusted Bob and a trusted Charlie and take the minimum to obtain the secret sharing rate.

Although the secret sharing rate must be minimised over the subsets, it is permitted (and indeed essential) to maximise the rate over the choice for which basis is used for key generation and which for certification. One might well ask, in the above calculation, why did we choose to have Bob make his guess of $X_A$ by measuring his $\hat{p}$ quadrature? For that matter, why did we choose to have Alice encode in $\hat{x}$? The answer to both lies in the correlation structure of the underlying graph state. This is nicely captured by the nullifiers of the graph state defined in Eq.~(\ref{nullifiers}), which allow us to read off Alice's optimal encoding choice and also Bob and Charlie's correct measurement given Alice's choice. Here the stabilisers are given by
\eqn{\mathbf{n} = \bk{\begin{array}{c}
     \hat{x}_A+\hat{p}_B-\hat{x}_C \\
     \hat{p}_A+\hat{x}_B \\
     \hat{p}_C-\hat{x}_B
\end{array}}.}
From this we can immediately read off that the optimal choice for Alice is to encode her key in $\hat{x}_A$, because it is correlated with variables of Bob and Charlie that are themselves \emph{not} correlated. To satisfy the structure of a secret sharing protocol, it must the the case that the key is encoded in a variable that is much more correlated with a \emph{collaborative}, joint variable of an authorised $k$-subset (here Bob and Charlie together) than it is for any $k-1$-subset (in this case, either Bob or Charlie individually). If Alice had chosen to encode in $\hat{p}_A$ then the second nullifier tells us her key would be well correlated with $\hat{x}_B$. The third nullifer tells us that, since $\hat{x}_B$ is correlated with $\hat{p}_C$ then $\hat{p}_A$ is correlated with $\hat{p}_C$ also. But this is precisely the problem! If $\hat{x}_B$ and $\hat{p}_C$ are correlated then the amount of information that Bob and Charlie have about $\hat{p}_A$ is almost identical to the amount either of them have individually, which makes secret sharing impossible. In contrast, because $\hat{p}_B$ and $\hat{x}_C$ are not correlated then a joint variable based on the combination of Bob and Charlie's measurements will be much more correlated with $\hat{x}_A$ than either would be individually which is exactly what we require. By the same logic if Bob were the dealer he should encode the secret in $\hat{p}_B$ and Charlie should use $\hat{x}_C$. Given an encoding choice the nullifiers also immediately define which of the variable players will be correlated with the key generation and check measurements of the dealer. 

Now that we know how to choose the key generation and certification measurements for a given choice of dealer, we turn to the question of the optimal choice of dealer. It turns out that for this case a positive secret sharing rate can be obtained for any choice of dealer. This is an attractive feature because, in principle, the states could be distributed and measured and the dealer chosen later on. However, assuming that the dealer is know beforehand, there asymmetry of the line graph means that there is an optimal choice. By inspection, for the three qubit line graph and symmetric transmission losses the correlations are identical for whoever is sent one of the two `ends' of the line graph. However, the participant who receives the middle node observes different correlations. These two possibilities were addressed in Fig.~\ref{asympcomp} from which we see that the optimal situation is where the dealer is given the middle node of the graph. Due the the alphabetic ordering convention, in our work the middle node is always given to Bob, so from now on make the optimal choice and designate Bob as the dealer. Finally, note that when the dealer is in the middle node then for, by construction, for the case considered here the secret sharing rate is identical regardless of which of the two `end' players (Alice or Charlie) is untrusted meaning the maximisation of the Holevo information in Eq.~(\ref{kasymp}) becomes redundant. However, if the dealer is an end node the correlations are not identical and the maximisation must be checked explicitly.

\begin{figure}[htb]
\includegraphics[width=0.45\textwidth]{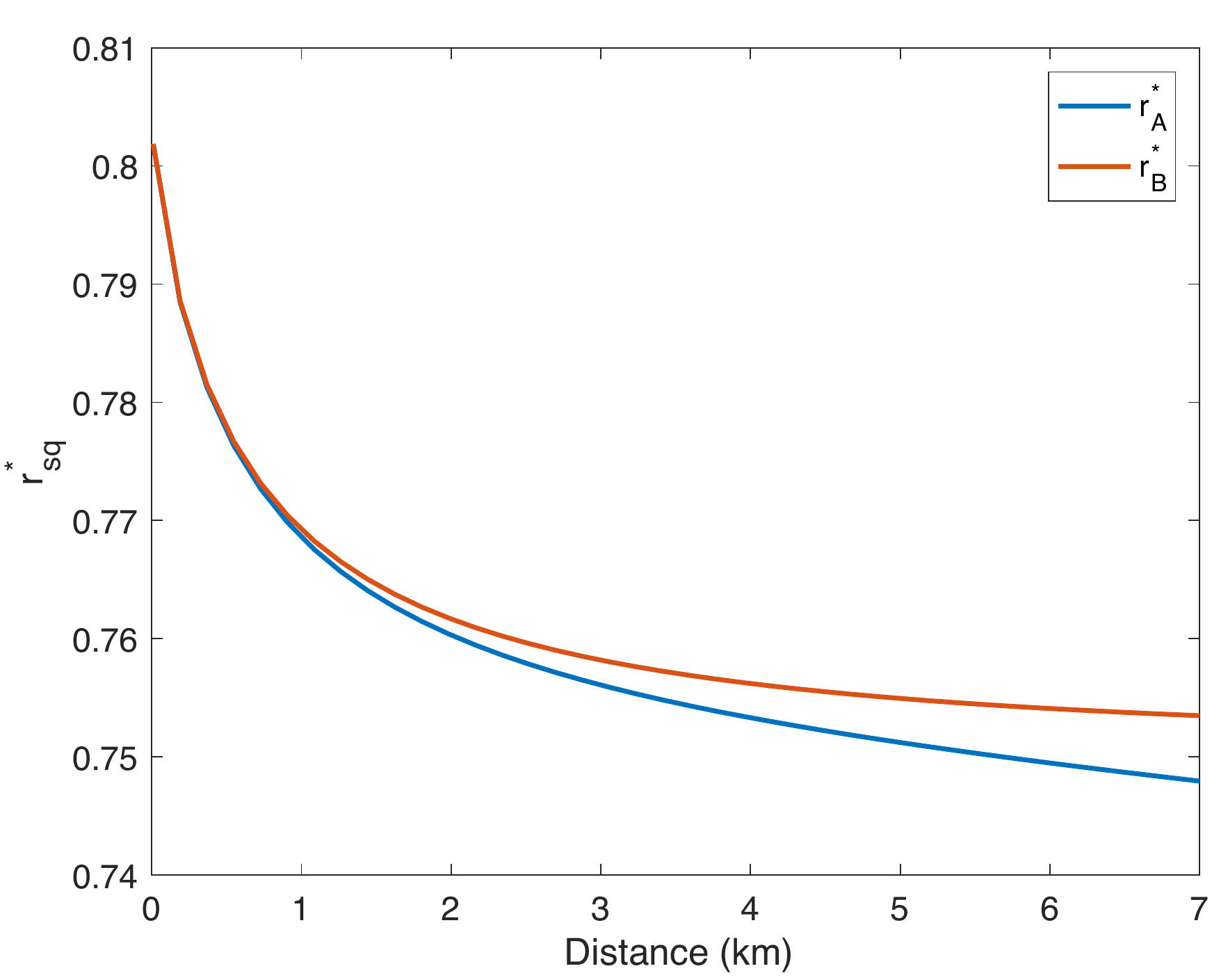}
\caption{Optimal value for the initial squeezing in the equivalent canonical graph state generation method given that the state is actually being created by offline squeezing via the Bloch-Messiah decomposition given in Eq.~(\ref{Gbloch}). The optimal values for the case where the middle node (red) or edge node (blue) are the dealer (corresponding to the curves in Fig.~\ref{asympcomp}) are plotted as a function of transmission distance. \label{ropt}}
\end{figure}

The final ingredient in evaluating the secret sharing rate is the allocation of squeezing resources. In order to make a fair, finite-squeezing comparison with a CVQKD protocol we will fix a maximum squeezing parameter $r_{\mathrm{\max}}$ that can be achieved. If we considered making the graph states via the canonical process of implementing CZ gates (which also require squeezing to implement) between already-squeezed states it is unclear how to easily constrain the CZ gain $g$ and the initial squeezing $r$ to satisfy our overall constraint. However, when using the Bloch-Messiah decomposition, which is in any case much more practical, this is straightforward. In this implementation, a canonical graph state of initial squeezing $r$ and CZ gain $g$ is decomposed into a set of single-mode squeezers with parameters $r_A, r_B$ and $r_C$ given by Eq.~(\ref{rbloch}). Now we simply constrain the largest of these to be less than $r_{\mathrm{max}}$ and then optimise Eq.~(\ref{kasymp}) over the achievable combinations of effective $r$ and $g$ for each transmission. The optimal choice of $0\leq r \leq r_{\mathrm{max}}$ for the curves in Fig.~\ref{asympcomp} is shown in Fig.~\ref{ropt}. We see that for the this case, with $r_{\mathrm{max}} \approx 1.76$, the optimal effective $r$ starts at just under half the maximum value and then declines with increasing losses before flattening out. Interestingly, there is a slightly different optimal choice depending upon whether the dealer is is the middle (Bob) or the edge (Alice and Charlie) of the graph.

Finally, we turn the the benchmark comparisons. For the PLOB bound we only need the effective transmission between Alice the players (this is the same since the network is symmetric) which is given by $T^2$. Substituting this in the result of Ref.~\cite{Pirandola:2017jk} and recalling that the rate must be halved since two network uses are required for a single secret sharing round  gives Eq.~(\ref{PLOB}). The CVQKD curves in Fig.~\ref{asympcomp} have been calculated by having the dealer create a two-mode squeezed vacuum state with squeezing given by $r_{\mathrm{max}}>0$, which is then transmitted through a lossy channel of transmission $T^2$ again modelled by beam-splitter mixing in a vacuum mode. The evolution for state generation and transmission between Alice Bob (also Charlie since the situation is symmetric) is given by
\eqn{\mathbf{N}_{ABV}^{\mathrm{QKD}} = \mathbf{BS}_{23}(T^2)\cdot\mathbf{BS}_{12}(1/2)\cdot\mathbf{S}_2(r_{\mathrm{max}})\cdot\mathbf{S}_1(r_{\mathrm{max}})\nn}
In Fig.~\ref{asympcomp}, we plot two comparison protocols, one where Alice makes a homodyne measurement (equivalently sends squeezed states) and one where she heterodynes (sends coherent states).  Explicitly we have,
\eqn{\chi(X_B:E) = S(AB) - S(A|x_B)}
for both protocols and,
\eqn{I(X_A:X_B) = \frac{1}{2}\log_2\bk{\frac{V_{X_B}}{V_{X_B|X_A}}}}
when Alice homodynes and
\eqn{I(X_A:X_B) = \frac{1}{2}\log_2\bk{\frac{2V_{X_B}}{V_{X_B|X_A}+1}}}
when Alice heterodynes and introduces an extra unit of shot noise. This CM allows the CVQKD rate of Eq.~(\ref{bss}) to be straightforwardly evaluated for an RR protocol where Bob switches between quadrature measurements. The conditional variances and symplectic eigenvalues to evaluate the above expressions can be obtained from Alice and Bob's modes of the CM $\mathbf{\Gamma} = \mathbf{N}_{ABV}^{\mathrm{QKD}}\cdot \mathbf{N}_{ABV\intercal}^{\mathrm{QKD}}$ and applying Eq.~(\ref{schur}) as necessary. Finally, the appropriate PLOB bound is given by Eq.~(\ref{PLOB}).

\subsection{Realistic, finite-size results \label{finres}}
Turning to evaluating the composable, finite-size secure fraction in Eq.~(\ref{secfrac}), we begin by fixing the target secrecy and correctness parameters ($\epsilon_s,\epsilon_c$). Based on our analysis of the asymptotic case we know to designate the middle node as the dealer and that because of the symmetry of the situation we will obtain the same secret fraction when assuming either player is dishonest. For each transmission we will use the optimal choice of the effective $r$ and $g$ found in the asymptotic case for graph state generation. There are several more parameters that can be chosen and (to some extent) optimised over: the detector resolutions ($\delta_{\mathbb{X}},\delta_{\mathbb{X}}$), the energy test beam-splitter $T_e$, the positive constants in the security proof ($\epsilon_1,\epsilon_\mu$) and the probability of any party choosing to measure in the key generation basis $p$. 

\begin{table}[ht]
    \centering
    \begin{tabular}{c|c|c}
      {\bf Symbol}   & {\bf Value} & {\bf Description} \\
      \hline
      $\epsilon_s$& $10^{-9}$& Secrecy parameter \\
      $\epsilon_c$& $10^{-9}$& Correctness parameter \\
        $\delta_{\mathbb{X}}$ & 0.1  & Key basis resolution\\
        $\delta_{\mathbb{P}}$ &  0.4 & Check basis resolution \\
        $M$ & 25 & Detector range\\
         $\epsilon_1$ &  $4\times10^{-11}$ & Security constant\\
         $\epsilon_\mu$ & $4\times10^{-20}$  & Security constant\\
         $T_e$ & 0.99 & Energy test beam-splitter \\
         $\alpha$& 28 & Energy test threshold\\
         $\beta$ & 0.98 & Reconciliation efficiency \cite{Pacher:2016te}\\
         \hline 
         
    \end{tabular}
    \caption{Parameters for realistic experimental model.}
    \label{tab:params}
\end{table}

Initial investigations showed that in the regimes of interest the keyrate is only weakly dependent on most of these parameters, except for the key generation probability $p$. This is because we are interested in parameters where the multi-partite and bi-partite schemes cross over, which is when both keyrates are still far above zero and with large block sizes $N=m+t$. In this regime, one can have a $p$ close to one whilst achieving sufficiently large parameter estimation measurements such that any statistical errors are small. This makes the parameters such as $\epsilon_1,\epsilon_\mu$ and $T_e$ less critical. Hence, in our analysis we chose fixed values for all parameters (Table~\ref{tab:params} except for $p$, which has been optimised over.

Given fixed values of all of the parameters, to evaluate the expected secret fraction we need only compute the expected value of the distance, the second moments, $V_A^{\mathrm{PE}},V_B^{\mathrm{PE}}$ and $V_d$ and the amount of reconciliation information $\ell_{\mathrm{EC}}$. Taking Alice to be the trusted parties carrying out parameter estimation we have,
\eqn{\EV{d(\mathbf{P}^\mathrm{PE}_A,\mathbf{P}^\mathrm{PE}_B)} \ee  \sum_{j_A,k_B=1}^{2M/\delta_{\mathbb{P}}} |j_A-k_B|\hspace{1mm}\mathrm{Pr}(j_A,k_B) \nn,\\
\EV{V_d} \ee  \sum_{j_A,k_B=1}^{2M/\delta_{\mathbb{P}}} |j_A-k_B|^2\hspace{1mm}\mathrm{Pr}(j_A,k_B) \nn, \\
\EV{V_A^{\mathrm{PE}}} \ee  \sum_{j_A=1}^{2M/\delta_{\mathbb{P}}} j_A^2\hspace{1mm}\mathrm{Pr}(j_A,k_B) \nn,\\
\EV{V_B^{\mathrm{PE}}} \ee  \sum_{k_B=1}^{2M/\delta_{\mathbb{P}}} k_B^2\hspace{1mm}\mathrm{Pr}(j_A,k_B), 
\label{moments}}
where the probability distribution for Alice and Bob's discretised measurement outcomes is given by
\eqn{\mathrm{Pr}(j_A,k_B) = \int_{I_{j}} dq_A\int_{I_{k}} dq_B \hspace{1mm}\mathrm{Pr}(q_A,q_B), \label{pjakb}}
where the integration intervals are defined in Eq.~(\ref{I}) and $\mathrm{Pr}(q_A,q_B)$ is the underlying distribution of Alice and Bob's parameter estimation variables. In the absence of an eavesdropper all the first moments vanish so we have dropped them. We have written $q_A$ and $q_B$ for the parameter estimation variables because, depending on the graph structure, either quadrature could have been designated for parameter estimation depending upon the initial graph structure. In this specific case, the parameter estimation observables are $\hat{p}_A$ and $\hat{x}_B$. A final point is that the players are free to scale their measurement results to avoid artificially underestimating their correlations. For example, in an asymmetric network where Charlie's channel is twice as lossy as the other players, it is clear that his measurement values will be correspondingly `damped' and all participants should take this into account. In this present case, given his knowledge of what the communication network should be in the absence of tampering, Bob can determine what scaling factor he should apply to maximise his correlations with Alice. Utilising Eq.~(\ref{rcond}) and the fact that prior to measurement all first moments should be zero, a measurement by Alice returning a value $p_A$ will project Bob's mode into a Gaussian state with mean vector
\eqn{\mathbf{r}_B \ee \mathbf{C}_{A,B}(\mathbf{P\Gamma}_A\mathbf{P})^{\mathrm{MP}}\text{diag}(p_A,0)\nn \\
&:=& \text{diag}(a \hspace{.5mm}p_A,0) \label{a}}
for a constant $a>0$. Therefore, if Alice re-scales her measurements by a factor of $a$, defining $q_A = a p_A$, then this new variable will be have a distribution centred about Bob's measurement value and a conditional variance of 
\begin{eqnarray}
V_{Q_A|X_B} &=& V_{Q_A} - \EV{Q_AX_B}^2/V_{X_B}\\ 
&=& a^2V_{P_A} - a^2\EV{P_AX_B}^2/V_{X_B} = a^2 V_{P_A|X_B}. \nonumber
\end{eqnarray}
This leads to a joint probability distribution for this re-scaled variable of.
\eqn{\mathrm{Pr}(q_A,x_B) \ee \mathrm{Pr}(x_B)\mathrm{Pr}(q_A|x_B) \nn\\
\ee \frac{e^{-\frac{\left(q_A-x_B\right)^2}{2 a^2V_{P_A|P_B}}-\frac{x_A^2}{2 V_{X_B}}}}{2 \pi  a \sqrt{V_{X_B} V_{P_A|X_B}}}.}
The expected correlations for a realistic implementation described by the CM given by Eq.~(\ref{GExp}). The quantity $V_{X_B}$ is given by the first entry of $\mathbf{\Gamma}_B$ and direct substitution in Eq.~(\ref{a}) gives
\eqn{a = \frac{g e^{2 r} \eta _c \eta _d \sqrt{\xi _A T_B}}{\eta _d \left(e^{2 r} T_B \eta _c-T_B \left(\xi _B+\eta _c\right)+\xi _B\right)+1}.}
Similarly, for the conditional variance by combining Eq.~(\ref{schur}) with Eq.~(\ref{GExp}) to obtain the conditional CM $\mathbf{\Gamma}_{A|x_B}$, the first entry of which gives
\begin{widetext}
\eqn{V_{P_A|X_B} \ee \frac{e^{-2 r}}{\eta _d \left(T_B \left(\left(e^{2 r}-1\right) \eta _c-\xi _B\right)+\xi _B\right)+1} \left [\left(\xi _B \eta _d+1\right) \left(\xi _A \eta _d \left(-e^{2 r} \left(\xi _A+\eta _c-1\right)+g^2 e^{4 r} \eta _c+\eta _c\right)+e^{2 r}\right) \right. \nn\\
 &+&  T_B \eta _d \left(e^{2 r} \left(\xi _B \left(\xi _A \eta _d \left(\xi _A+\eta _c-1\right)-1\right)+\eta _c \left(\xi _A \eta _d \left(\xi _A+2 \eta _c-1\right)-1\right)\right)-\xi _A \eta _c \eta _d \left(\xi _B+\eta _c\right)\right) \nn \\
 &-&\left. T_B \eta _d e^{4 r} \eta _c \left(\xi _A \eta _d \left(\xi _A+g^2 \xi _B+\left(g^2+1\right) \eta _c-1\right)-1\right) \right]. }
 
\end{widetext}
With an explicit form for $(q_A,x_B)\mapsto\mathrm{Pr}(q_A,x_B)$, Eq.~(\ref{pjakb}) can be numerically integrated and the quantities in Eq.~(\ref{moments}) computed. 

\begin{figure}[htb]
\includegraphics[width=0.45\textwidth]{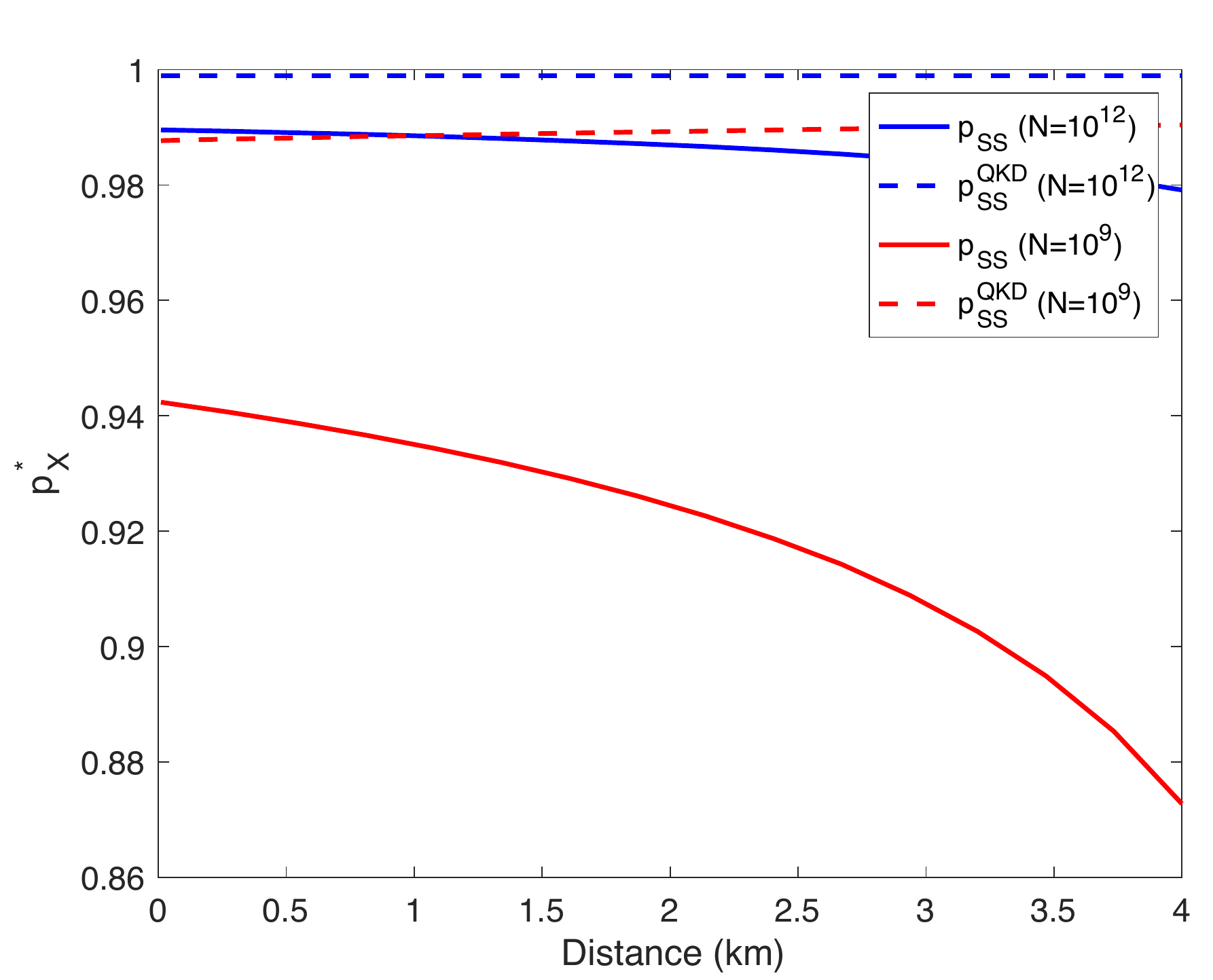}
\caption{Optimal value for the probability of key generation as a function of transmission distance for multi-partite (solid) or bi-partite (dashed) protocols. Optimal curves for block sizes of $m=10^{12}$ (blue) and $m=10^9$ (red) are shown.  \label{popt}}
\end{figure}

The last ingredient to evaluate the secret fraction are the block sizes involved and the expected number of network uses required to achieve them. We proceed by first fixing the desired block size for the key generation $m$. For a given value of $p$ we expect this to take,
\eqn{\EV{L} = \frac{m}{p^3}}
network uses. The expected number of parameter estimation rounds with any player should then be given by 
\eqn{\EV{t_j} = (1-p)^2L. \label{mt3}}
This is everything required to compute the smooth min-entropy and finally the information leakage during error correction is well approximated by \cite{Furrer:2014eq,Pacher:2016te},
\begin{equation}{\ell_{\mathrm{EC}} =  m\bk{H(X_B) - \log_2(\delta_{\mathbb{X}})- \beta I(X_A:X_B)}}
\end{equation}
where $H(X_B)$ and $I(X_A:X_B)$ are the entropy and mutual information of the ideal Gaussian distributed variables. 
This gives
\begin{equation}
\ell_{\mathrm{EC}} = \frac{m}{2}\bk{\log_2\bk{\frac{2\pi e V_{P_B}}{\delta_{\mathbb{X}}}}- \beta \log_2\bk{\frac{V_{P_B}}{V_{P_B|X_A,X_C}} }}
\label{lECGauss}
\end{equation}
where we are abusing notation slightly by continuing to use $\delta_{\mathbb{X}}$ to refer to the resolution in the key generation basis, while explicitly using the fact that the key generation basis is actually made by Bob measuring in the $\hat{p}$ quadrature. The necessary variances and conditional variances are again given by Eq.~(\ref{GExp}). For each transmission the expected secret fraction can now be computed for the fixed parameters of Table~\ref{tab:params} and an optimised value of $p$ which is shown in Fig.~\ref{popt}.

Turning to our benchmarks, a fair comparison with a bi-partite QKD protocol is obtained by considering an bi-partite CVQKD protocol implemented with the same escape, detector and fibre coupling efficiencies and maximum allowed squeezing of Table \ref{tab:exp}. However, this bQSS scheme will only require two squeezers to make an EPR state rather than the three required for the graph state. The second difference is that we must now consider transmission through the two `arms' of the network to connect Bob with Alice and Charlie in turn. To maximise the key rate in this situation it is optimal to carry out a reverse reconciliation (RR) protocol \cite{Grosshans:2002p5549}, where the players (Alice and Charlie) transmit a quantum state to the dealer (Bob) and then try and guess his measurement outcome. Furthermore, previous work on CVQKD with `entanglement-in-the-middle' has established that this is always inferior to a standard RR protocol \cite{Weedbrook:2013gn}. In our language, for carrying out CVQKD, the Hub-Out strategy is always inferior. 

A realistic protocol with this transmission strategy from Alice to Bob is described by an overall evolution of a 7 mode system (a mode for Alice and Bob, an extra mode, $B_e$, for Bob's energy test, two thermal modes, $E_A,E_B$, for the transmission through the thermal loss network from Alice to the hub and from the hub to Bob and 4 vacuum modes for the detector and experimental efficiencies for each of Bob and Alice) given by
\eqn{\mathbf{N}_{\mathrm{exp}}^{\mathrm{QKD}} \ee \mathbf{BS}_{B,V_3}(\eta_d)\cdot\mathbf{BS}_{A,V_4}(\eta_d)\cdot\mathbf{BS}_{B,B_e}(T_e)\nn \\
&\times&\mathbf{BS}_{B,E_B}(T) \cdot\mathbf{BS}_{B,E_A}(T)\cdot\mathbf{BS}_{B,V_2}(\eta_c)\nn \\
&\times&\mathbf{BS}_{A,V_1}(\eta_s)\cdot\mathbf{BS}_{A,B}(1/2)\cdot\mathbf{S}_A(r_{\mathrm{max}})\nn \\
&\times&\mathbf{S}_B(r_{\mathrm{max}}).}
Note that this bi-partite implementation experiences slightly less loss than the multi-partite version because Alice's mode can be directly detected rather than being coupled into an optical fibre. This is why Bob's mode experiences an efficiency of $\eta_c = \eta_f\eta_s$ whereas Alice's experimental efficiency is $\eta_s$.

 Next, we must calculate the block sizes for a given key generation probability, $p$, which will be optimised over. Here, we see the importance of a finite-size analysis since the QKD protocol has a different, and strictly higher, performance. This is because, in a bi-partite scheme, Bob can take his partner to be trusted and so they can agree ahead of time on a random selection of runs to be used for parameter estimation. This means, for a fixed $p$ and desired $m$, the total number parameter estimation strings is given by
\eqn{\EV{L} = \frac{m}{p},\hs \EV{t} = (1-p)L \label{mtqkd}.}
In this sense, the QKD scheme is always more efficient as there are not wasted rounds due to basis mismatch. The penalty for this is that the participants will need to refresh the extra pool of pre-shared key of length $h_2(p)L$ bits to use for choosing the parameter estimation rounds in the next run of the protocol. This means the length of secret key for the CVQKD protocol is given by
\eqn{\ell_{QKD} \ee \frac{1}{2} \left( \hmin^\epsilon(\bm{X}^m_B|E) - \ell_{\mathrm{EC}} - \log_2\frac{1}{\epsilon_c\epsilon_1^2} \right. \nn \\
&+&   2 - h_2(p)L \bigg ).}
With the protocol parameters fixed to be the same as in Table~\ref{tab:params} and given target $m$, the block sizes are given by Eq.~(\ref{mtqkd}) and all the necessary correlations to compute the expected values of the parameter estimation quantities in Eq.~(\ref{moments}) can be obtained from the first two modes of the global CM $\mathbf{\Gamma} = \mathbf{N}_{\mathrm{exp}}^{\mathrm{QKD}}\cdot\mathbf{N}_{\mathrm{exp}}^{\mathrm{QKD\intercal}}$. These are then used to lower bound the min-entropy via Eq.~(\ref{secfracapp}) and the information leakage in in Eq.~(\ref{lECGauss}) which gives the secret fraction. To make the curves in Fig.~\ref{finitecomp} the key generation probability has been optimised over and the resulting optimal probability for the CVQKD protocol is shown in Fig.~\ref{popt}.

Finally, to make a fair comparison with the PLOB bound, we set the total transmission to include efficiency of the fibre-coupling. This means the asymptotic PLOB bound becomes
\eqn{K_{\mathrm{PLOB}} \ee -\frac{1}{2}\log_2 \bk{1-\eta_f T^2}.}  

\section{Discussion of the Williams et al. protocol\label{Williams}}

In this section, 
we further discuss the alternative security proof method of Williams et al. \cite{Williams:2019kb} for a variant of the original HBB protocol. Here the multi-partite state is a DV GHZ state and the players switch between Pauli $X$ and $Y$ measurements. In this work, the parameter estimation checks now always involve all players, which is precisely how participant attacks entered in the first place. However, in the protocol 
of Ref.~\cite{Williams:2019kb} they are thwarted by the introduction of a randomisation in the order of announcements. After all states have been transmitted, the dealer randomly chooses a round to be disclosed for parameter estimation and and randomly picks an untrusted subset to announce first. As long as the players have no way to know which rounds will be used for parameter estimation and whether or not they will have to announce first then there is no way for dishonest players to meddle with the statistics.

Since it is crucial that the players cannot tell \emph{a priori} whether a round will eventually be used for parameter estimation or key generation, it is essential that there is no pre-designated key basis and that basis choices are made with $p=1/2$. To see why, consider the case of two players where it has been pre-designated that the dealers $Y$ measurement would be the key and their $X$ basis used as the check. Now, even if a dishonest player is told to announce their basis first, the mere fact that it is a parameter estimation round means they still cheat the test because they know the dealer will be measuring in the $X$ basis. Assuming they have made the same attack as in Ref.~\cite{Karlsson:1999ez}, they now have both particles of the original GHZ state and also one half of an entangled pair shared with the honest player. First, they can learn the dealers measurement perfectly by measuring their GHZ particles. Secondly, the dishonest player can announce either $X$ (respectively $Y$) and then also measure their Bell pair shared with the honest player in that basis. The round will now only be kept if the honest player also measured $X$ (respectively $Y$). If it is not discarded, the dishonest player knows what values the dealer and honest player obtained, and can thus announce the correct value themselves.

This is why, in Ref.~\cite{Williams:2019kb}, it is stipulated bases are chosen with $p=1/2$ and that a portion of each basis is used to certify key generated in the other. As explained in the main text, this means the protocol cannot be made arbitrarily efficient. Whilst the factor of 1/2 removes any possible advantage in a ($2,2$)-threshold scenario, for larger $n$ it is possible that this strategy could be effective and possibly even superior. Firstly, the probability that a round is useful remains at 1/2 for arbitrarily many players. Secondly, because all players are involved in the check measurements the correlations observed will be higher than in the strategy pursued here where in a given check the untrusted subset is effectively traced out.

However, the necessary amount of parameter estimation data now scales in the number of players. Each scenario of a given untrusted subset announcing first must be treated as its own QKD protocol. This means for a $(n,k)$-scheme is will be necessary to acquire $\binom{n}{k}$ many parameter estimation data sets. A detailed analysis of the best strategy for a given finite block size and larger number of players is an important question for future work.

\end{appendix}

%

\end{document}